\documentclass[sigconf, nonacm]{acmart}

\usepackage{xspace}

\newcommand\vldbdoi{XX.XX/XXX.XX}
\newcommand\vldbpages{XXX-XXX}
\newcommand\vldbvolume{14}
\newcommand\vldbissue{1}
\newcommand\vldbyear{2020}
\newcommand\vldbauthors{\authors}
\newcommand\vldbtitle{\shorttitle} 
\newcommand\vldbavailabilityurl{https://github.com/curtis-sun/LLM4Rewrite}
\newcommand\vldbpagestyle{empty}

\usepackage{graphicx}
\usepackage{balance}  
\usepackage{epsfig}
\usepackage{epstopdf}
\usepackage{latexsym}
\usepackage{enumitem}
\PassOptionsToPackage{noend}{algpseudocode}
\usepackage{algpseudocode}
\usepackage{pifont}
\usepackage{epsfig}
\usepackage{amssymb}
\usepackage{amsmath}
\usepackage{amsfonts}
\usepackage{stmaryrd}
\usepackage{url}
\usepackage{multirow}
\usepackage{array}
\usepackage[normalem]{ulem}
\usepackage{color}
\usepackage{cleveref}
\usepackage{xspace}
\usepackage{mathtools}
\usepackage{soul}
\usepackage{listings}
\usepackage{xcolor}
\usepackage{tikz}
\newsavebox{\blackball}
\newsavebox{\greenball}

\usepackage[utf8]{inputenc}
\usepackage{amsmath}

\usepackage{bbding}
\usepackage{amssymb}
\usepackage{epstopdf}
\usepackage{epsfig,endnotes}
\usepackage{url}
\usepackage{array}
\usepackage{booktabs}
\usepackage{threeparttable}
\usepackage{tabulary}
\usepackage{CJKutf8}
\usepackage[most,breakable]{tcolorbox}
\usepackage{etoolbox}
\usepackage{fancyhdr}
\usepackage{lipsum}
\usepackage[fencedCode]{markdown}
\usepackage{tikz}
\usepackage{hyperref}
\usepackage{balance}  
\usepackage{tikz}
\usepackage{lipsum} 
\usepackage{marginnote}
\usepackage{makecell}

\usepackage{subcaption}

\usepackage[lined,boxed,vlined,ruled,linesnumbered]{algorithm2e}

\definecolor{shadecolor}{rgb}{0.92,0.92,0.92}
\usepackage{listings,upquote,soul,framed}
\colorlet{shadecolor}{gray!20}

\newcommand{\hi}[1]{\vspace{.25em} \noindent {\bf #1}\xspace}

\newcommand{\llm}{\textsc{LLM}\xspace}
\newcommand{\llms}{\textsc{LLMs}\xspace}
\newcommand{\QA}{Q\&A\xspace}
\newcommand{\QAs}{Q\&As\xspace}

\newcommand{\oursys}{\textit{R-Bot}\xspace}
\newcommand{\rbotfour}{\emph{R-Bot (GPT-4)}\xspace}
\newcommand{\rbotthree}{\emph{R-Bot (GPT-3.5)}\xspace}

\newcommand{\lr}{\textit{LearnedRewrite}\xspace}
\newcommand{\vanillafour}{\emph{GPT-4}\xspace}
\newcommand{\vanillathree}{\emph{GPT-3.5}\xspace}

\begin{document}

\title{R-Bot: An LLM-based Query Rewrite System}

\author{Zhaoyan Sun}
\affiliation{%
  \institution{Tsinghua University}
}
\email{szy22@mails.tsinghua.edu.cn}

\author{Xuanhe Zhou}
\affiliation{%
  \institution{Shanghai Jiao Tong University}
}
\email{zhouxh@cs.sjtu.edu.cn}

\author{Guoliang Li}
\affiliation{%
  \institution{Tsinghua University}
}
\email{liguoliang@tsinghua.edu.cn}

\author{Xiang Yu}
\affiliation{%
  \institution{Huawei Company}
}
\email{yuxiang44@huawei.com}

\author{Jianhua Feng}
\affiliation{%
  \institution{Tsinghua University}
}
\email{fengjh@tsinghua.edu.cn}

\author{Yong Zhang}
\affiliation{%
  \institution{Tsinghua University}
}
\email{zhangyong05@tsinghua.edu.cn}


\begin{abstract}
Query rewrite is essential for optimizing SQL queries to improve their execution efficiency without changing their results. Traditionally, this task has been tackled through heuristic and learning-based methods, each with its limitations in terms of inferior quality and low robustness. Recent advancements in \llms offer a new paradigm by leveraging their superior natural language and code comprehension abilities. Despite their potential, directly applying \llms like GPT-4 has faced challenges due to problems such as hallucinations, where the model might generate inaccurate or irrelevant results. To address this, we propose \oursys, an \llm-based query rewrite system with a systematic approach. We first design a multi-source rewrite evidence preparation pipeline to generate query rewrite evidences for guiding \llms to avoid hallucinations. We then propose a hybrid structure-semantics retrieval method that combines structural and semantic analysis to retrieve the most relevant rewrite evidences for effectively answering an online query. We next propose a step-by-step \llm rewrite method that iteratively leverages the retrieved evidences to select and arrange rewrite rules with self-reflection. We conduct comprehensive experiments on real-world datasets and widely used benchmarks, and demonstrate the superior performance of our system, \oursys, surpassing  state-of-the-art query rewrite methods. The \oursys system has been deployed at Huawei and with real customers, and the results show that the proposed \oursys system achieves lower query latency. 
\end{abstract}

\maketitle                                                                                                                                                  
\pagestyle{\vldbpagestyle}
\begingroup\small\noindent\raggedright\textbf{PVLDB Reference Format:}\\
\vldbauthors. \vldbtitle. PVLDB, \vldbvolume(\vldbissue): \vldbpages, \vldbyear.\\
\href{https://doi.org/\vldbdoi}{doi:\vldbdoi}
\endgroup
\begingroup
\renewcommand\thefootnote{}\footnote{\noindent
This work is licensed under the Creative Commons BY-NC-ND 4.0 International License. Visit \url{https://creativecommons.org/licenses/by-nc-nd/4.0/} to view a copy of this license. For any use beyond those covered by this license, obtain permission by emailing \href{mailto:info@vldb.org}{info@vldb.org}. Copyright is held by the owner/author(s). Publication rights licensed to the VLDB Endowment. \\
\raggedright Proceedings of the VLDB Endowment, Vol. \vldbvolume, No. \vldbissue\ %
ISSN 2150-8097. \\
\href{https://doi.org/\vldbdoi}{doi:\vldbdoi} \\
}\addtocounter{footnote}{-1}\endgroup

\ifdefempty{\vldbavailabilityurl}{}{
\vspace{.3cm}
\begingroup\small\noindent\raggedright\textbf{PVLDB Artifact Availability:}\\
The source code, data, and/or other artifacts have been made available at \url{\vldbavailabilityurl}.
\endgroup
}

\vspace{-.5em}
\section{Introduction}
\label{sec:intro}
\vspace{-.25em}

Query rewrite is designed to transform an SQL query into a logically equivalent version that is more efficient to execute, playing a crucial role in enhancing query performance in numerous practical scenarios. In Huawei's real-world database deployments, slow queries are regularly identified and optimized for improved performance. For instance, while migrating an enterprise's core application, we rewrote 20 critical queries, leading to a 3.7x reduction in workload latency. 
Despite its significance, the process of query rewrite is NP-hard~\cite{zhou2021learned,zhou2023learned}, meaning there is a vast collection of possible rewrite rules, and the number of potential rule combinations increases exponentially. This complexity makes identifying an effective combination of rules a challenging and laborious task. There are two main paradigms for addressing this challenge.

\begin{table}[!t]
\centering
\caption{\oursys v.s. Existing Query Rewrite Methods.}\vspace{-1em}
\label{tab:rule-rewrites}
\hspace*{-1em}\begin{tabular}{|c|c|c|c|c|}
\hline
Method  & \makecell[c]{Heuristic \\ Fixed Order}                             & \makecell[c]{Heuristic \\ Exploring }                           & \makecell[c]{Traditional \\ Learning}                           & \makecell[c]{\oursys \\ LLM\&Evidence}                        \\ \hline
Train                     & No                                & No                              & \textit{Yes}                              & No                           \\ \hline
\#Rules            & High                             & \textit{Low}                            & High                           & High                        \\ \hline
Quality                      & \textit{Low}                              & \textit{Low}                           & High                           & High                        \\ \hline
Robust                      & {High}                              & {High}                           & \textit{Low}                           & High                        \\ \hline
\end{tabular}
\vspace{-1.75em}
\end{table}

\hi{Heuristic-based Methods.} Some heuristic-based methods apply the rules in a fixed order derived from practical experience (e.g., PostgreSQL~\cite{postgresql}).  However, they may not achieve optimal results for queries requiring different rule orders, thereby risking the omission of essential rewrite sequences. Furthermore, other heuristic-based methods (e.g., Volcano~\cite{graefe1993volcano}) attempt to comprehensively explore various rule orders through heuristic acceleration. Nevertheless, they might overlook dependencies among rules, such as an initially inapplicable rule could be activated by another, leading to the potential neglect of vital rewrite sequences. Thus, heuristic-based approaches are often criticized for their {\it inferior quality}.

\hi{Learning-based Methods.}
To optimize query rewrite, learning-based methods have been proposed~\cite{zhou2021learned, zhou2023learned}. These methods employ neural networks that are trained on historical query rewrites to identify and apply the most advantageous rules for rewriting a query. However, learning-based approaches face criticism for their {\it low robustness}. For example, models trained through these methods struggle to adapt to unseen database schemas without undergoing additional training on {new query rewrite examples (e.g., hundreds of examples)}, which may not be readily available in real scenarios.

Recent advances in large language models (\llms) have shown superiority in understanding natural language and code, as well as reasoning ability~\cite{hu2024minicpm,achiam2023gpt,du2024evaluating,imani2023mathprompter,zheng2024revolutionizing,liu2024survey,fan2024combining,li2024dawn,zhou2025cracksql,li2025data+,wang2025idatalake,wang2025aop,zhou2025cracking,sun2025data}. As \llms can capture the query rewrite capabilities by pre-training from database forums and codes, encompassing both the direct rewriting ability of applying a rule and the indirect rewriting ability to draw inspiration from multiple dependent rules, we can leverage \llms to guide the query rewrite, particularly for slow queries that often remain as bottlenecks.

To realize this {target}, we aim to develop an {\it \llm-based query rewrite system} with {three} main advantages: 
\emph{\textit{(1) High Quality.}} On one hand, our system can figure out the potential rules that have implicit relations with the query (e.g., ones that become applicable only after applying other rules). On the other hand, our system can understand the interrelations among rules and generate an effective sequence of rules for holistic improvements. \emph{\textit{(2) Zero-Shot Robustness.}} Unlike conventional learning-based approaches that are limited to in-distribution data~\cite{zhou2021learned,zhou2023learned}, our system harnesses \llms, whose extensive pre-training empowers it to adapt to new datasets seamlessly without the need for additional retraining. \emph{\textit{(3) Executability and Equivalence.}} Our system ensures the rewritten query is both executable and functionally equivalent to the original one, as it performs query rewrite by selecting and ordering well-crafted rewrite rules from established query optimization engines, rather than directly rewriting the queries via \llms.

However, directly utilizing \llms for query rewrite proves to be ineffective due to their tendency for hallucination~\cite{huang2023survey}. For example, despite being pre-trained on a vast corpus of query rewrite data (e.g., Stack Overflow~\cite{stackoverflow}), employing the advanced \llm GPT-4 to directly rewrite queries in DSB benchmark~\cite{ding2021dsb} yielded only a 5.3\% success rate, which is significantly low. This highlights two primary challenges associated with leveraging \llm for query rewrite.

\hi{C1: How to mitigate \llm's factuality hallucination in query rewrite?} It's common for \llm to encounter confusion during query rewrite, suggesting an intuitive approach of guiding \llm with specific rewrite evidences (e.g.,  database \QAs, database manuals and codes, forum, etc) closely related to the query. However, several challenges arise from this approach. Firstly, \llm often struggles to interpret the raw rewrite evidence due to difficulties in aggregating fragmented knowledge across various rewrite documents and understanding complex query rewrite codes. Secondly, it's crucial to sift through and identify the most beneficial rewrite evidences to serve as references for \llm, thereby steering it towards a more efficient rule selection for query rewrite.

\hi{C2: How to mitigate \llm's faithfulness hallucination in query rewrite?} \llm  encounters challenges in accurately analyzing complex queries, such as those containing multiple sub-queries, and in fully leveraging detailed rewrite evidences. Thus, it becomes essential to develop a multi-step \llm rewrite method that breaks down the query rewrite process into more manageable segments, thereby aligning with \llm's capabilities for better performance.

\begin{table*}[!t]
\centering
\vspace{-2em}
\caption{Example Rewrite Rules.}\vspace{-1em}
\label{tab:rewrite-rules}
\begin{tabular}{|l|l|l|l|l|}
\hline
   & Rule                        & Condition & Transformation & Matching Function \\ \hline
$r_1$ & FILTER\_SUB\_QUERY\_TO\_JOIN        & \begin{tabular}[l]{@{}l@{}}Scalar, ``IN'', or ``EXISTS'' sub- \\ query in ``WHERE'' clause.\end{tabular}
  & \begin{tabular}[l]{@{}l@{}}Transformed to join on\\ the correlated column.\end{tabular} & \begin{tabular}[l]{@{}l@{}}b -> b.operand(Filter.class)\\ .predicate(containsSubQuery);\end{tabular} \\ \hline
$r_2$ & FILTER\_INTO\_JOIN &
\begin{tabular}[l]{@{}l@{}}Filter condition with column\\ from one join side.\end{tabular}
  & \begin{tabular}[l]{@{}l@{}}Push down condition to\\ filter non-nullable side.\end{tabular} & \begin{tabular}[l]{@{}l@{}}b -> b.operand(Filter.class)\\ .oneInput(Join.class);\end{tabular} \\ \hline  
$r_3$ & AGGREGATE\_PULL\_UP\_CONSTANTS &
\begin{tabular}[l]{@{}l@{}}Some ``GROUP BY'' key is\\ constant across rows.\end{tabular}
  & \begin{tabular}[l]{@{}l@{}}Remove constant key.\\ Constant may project.\end{tabular} & \begin{tabular}[l]{@{}l@{}}b -> b.operand(Aggregate.class)\\ .predicate(hasConstantExps);\end{tabular} \\ \hline 
\end{tabular}\vspace{-1.25em}
\end{table*}

To tackle the above challenges,  we propose \oursys, an \llm-based query rewrite system designed with a systematic approach. First, we gather and prepare rewrite evidences from diverse sources, including integrated {well-formatted} rewrite rules aggregating from rewrite documents and summarizing from the complex rule codes, as well as high-quality \QAs from database forums (addressing \textbf{C1}). Second, for an input SQL query, we propose a hybrid structure-semantics method for retrieving pertinent evidences, including rewrite rules by matching functions and rewrite \QAs with both query structure and rewrite semantics similarities (addressing \textbf{C1}). To enable \llm to comprehend the rewrite evidences, we synthesize this information into {\it rewrite recipes}, which detail in natural language how to utilize  the \QAs and rewrite rules for rewriting an SQL query, and we retrieve most relevant rewrite recipes to guide \llms for query rewrite (addressing \textbf{C2}).  Third, we design a step-by-step \llm rewrite algorithm, which guides \llm to iteratively utilize the rewrite recipes to refine its rewrite rule selection and ordering, possibly choosing a more promising rule sequence by reflecting and self-improving the rewrite process (addressing \textbf{C2}).

\hi{Contributions.} In summary, we make the following contributions.

\noindent $(1)$ We develop an \llm-based query rewrite system, \oursys, which can select an effective rewrite rule sequence to guide query rewrite engines to rewrite a query (see Section ~\ref{sec:overview}).

\noindent $(2)$ We design a multi-source rewrite evidence preparation pipeline, including clustering-based document reorganization and hierarchical code summarization for rewrite codes (see Section \ref{sec:evidence-prepare}).

\noindent $(3)$ We propose a hybrid structure-semantics retrieval method for retrieving relevant rewrite evidences (see Section \ref{sec:evidence-retrieval}).

\noindent $(4)$ We propose a step-by-step \llm rewrite method that iteratively leverages the retrieved {\QAs and} rewrite recipes to select and arrange rewrite rules with self-reflection (see Section \ref{sec:llm-rewrite}).

\noindent $(5)$ Our experimental results on real-world datasets and widely used benchmarks demonstrate that \oursys can significantly outperform existing state-of-the-art query rewrite methods. We have deployed \oursys at Huawei with real customers, achieving much higher real-world query rewrite performance (see Section \ref{sec:experiments}).

\vspace{-.5em}
\section{Preliminaries}\label{sec:pre}
\vspace{-.25em}
\subsection{Query Rewrite}\label{sec:rule-query-rewrite}
\vspace{-.25em}

Rewriting a query involves numerous query transformations, and we typically identify and summarize common transformation patterns into rewrite rules. For instance, a query rewrite rule could involve replacing an outer join with an inner join when they are equivalent. The composition of these rules offers the flexibility to accommodate a wide range of query rewrite requirements.

\vspace{-.5em}
\begin{definition}[Rewrite Rule]\label{def:rewrite-rule}
A rewrite rule $r$ is denoted as a triplet $(c,t, f)$, where $c$ is the condition to use the rule,  $t$ is the transformation to be applied, and $f$ is a matching function used for evaluation. If a query $q$ satisfies the condition $c$ as determined by the matching function $f$, then the transformation $t$ can be applied to the query $q$, resulting an equivalent rewritten query $q^{[r]}$.
\end{definition}
\vspace{-.5em}

For example, \autoref{tab:rewrite-rules} shows some query rewrite rules. Since the column ``comm'' with condition ``comm=100'' in the SQL query satisfies the condition ``some `GROUP BY' key is constant across rows'' of rule $r_3$, we can apply the rule and remove the column ``comm'' from the ``GROUP BY'' clause.

Numerous rewrite rules have been pragmatically incorporated into database products, as evidenced in existing literature and products such as PostgreSQL~\cite{postgresql}, MySQL~\cite{mysql}, Apache Calcite~\cite{begoli2018apache}.
However, when applying the rules to rewrite a query, it is often cumbersome to decide the best rule sequence for two primary reasons.
First, since a rewrite rule sometimes degrades the query (e.g., $q^{[r]}$ with higher execution latency than $q$), we should examine whether or not to use the rule. Second, it is also important to decide the order of applying the rules. For instance, applying one rule may render another rule obsolete. Thus query rewrite aims to find an optimal rule sequence to rewrite a query in order to minimize the execution cost of the rewritten query, which is formulated as below.

\vspace{-.5em}
\begin{definition}[Rule-based Query Rewrite]\label{def:rule-rewrite}
Consider a query $q$ and a set of rewrite rules $R$. Assume $\alpha$ is a sequence of certain rewrite rules selected from $R$. Let $q^\alpha$ denote the rewritten query by sequentially applying the rules in $\alpha$ to rewrite $q$. Query rewrite aims to obtain an optimal rule sequence $\alpha^*$, such that the execution cost of $q^{\alpha^*}$ is minimized among all possible rewritten queries of $q$.
\end{definition}
\vspace{-.5em}

The query rewrite problem has been proven to be NP-hard~\cite{zhou2021learned}. Traditional methods cannot select high-quality rules. To address this limitation, we advocate for utilizing \llm to select rewrite rules. To address the hallucination problem of \llms~\cite{huang2023survey}, we perform an offline stage to extract query rewrite evidences, including rewrite \QAs and rewrite rule specifications, and store them as \QA repository and rewrite rule specification repository respectively. During the online phase, given a SQL query, we retrieve relevant \QAs and rule specifications, and generate rewrite recipes, which outline how to rewrite a query using rewrite \QAs and rule specifications, assisting \llms in comprehending the rewrite evidences. With the assistance of the rewrite recipe, \llms are then guided through a step-by-step process to judiciously select and apply rewrite rules to the query. Next we formally define these notations. 

\begin{figure}[!t]
	\centering
\includegraphics[width=1.07\linewidth]{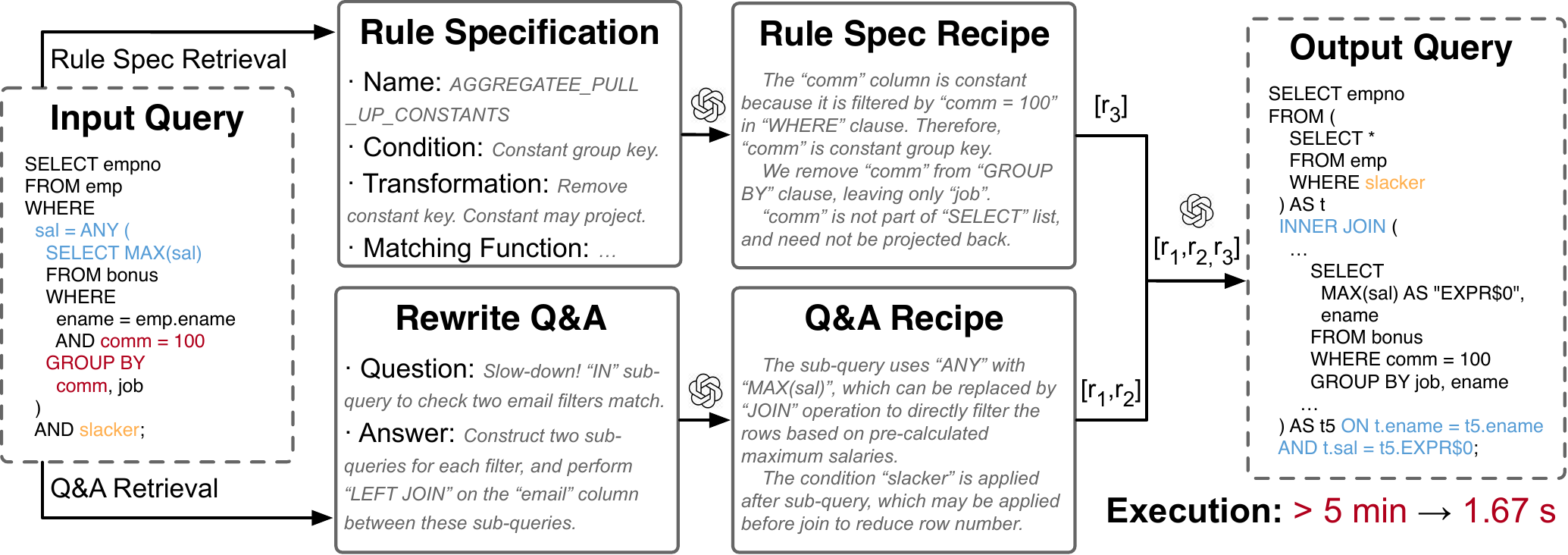}
 \vspace{-2.25em}
	\caption{Query Rewrite Example: the sequence of rewrite rules $[r_1,r_2,r_3]$ is found by \oursys based on retrieved rewrite rule specifications and rewrite \QAs.}
	\label{fig:query-rewrite}
 \vspace{-1.25em}
\end{figure}

\begin{table}[!t]
\centering
\caption{{Rewrite Rule Specification v.s. Rewrite Rule.}}
\vspace{-1em}
\label{tab:rewrite-rule-spec}
\begin{tabular}{|c |c |c |}
\hline               & Rule Specification      & Rewrite Rule                   \\ \hline Mapping       & Single/Multiple Rules           & Single Rule                            \\ \hline  Source        & Document/Code           & Code in Single Engine                    \\ \hline Clarity & \llm-Readable             & Compiler-Readable                            \\ \hline Selection     & Match Function & LLM\&Match Function \\ \hline
\end{tabular} \vspace{-1.75em}
\end{table}

\vspace{-.5em}
\begin{definition}[Rewrite Q\&A]\label{def:rewrite-QA}
A rewrite \QA includes a query rewrite question and a rewrite answer on how to rewrite the query in natural language. 
\end{definition}

\vspace{-1em}
\begin{definition}[Rewrite Rule Specification]\label{def:rewrite-specif}
A rewrite rule specification is used to describe a rewrite rule $(c, t, f)$ using natural language, which is also a triplet $(nc, nt, f)$, where $nc$ describes the condition and $nt$ describes how to apply the transformation in natural language. 
\end{definition}
\vspace{-.5em}

For example, \autoref{fig:query-rewrite} shows a rewrite rule specification derived from Calcite code of rule ``AGGREGATE\_PULL\_UP\_CONSTANTS'', which explains the condition and transformation in natural language. Besides, the rule specification will be retrieved as evidence in this example, and we can use this evidence to rewrite the query.

Based on the general concepts, we further introduce definitions related to particular query rewrite, including rewrite \QA recipe and rewrite rule specification recipe.

\vspace{-.5em}
\begin{definition}[Rewrite \QA Recipe]\label{def:QA-recipe}
Given a query $q$ and a \QA, a \QA recipe provides instructions in natural language on how to utilize the \QA to rewrite the query $q$. 
\end{definition}

\vspace{-1em}
\begin{definition}[Rewrite Rule Specification Recipe]\label{def:specification-recipe}
Given a query $q$ and a rewrite rule specification, a rewrite rule specification recipe describes how to use the rewrite rule specification to rewrite the query $q$ in natural language. 
\end{definition}
\vspace{-.5em}

For instance, \autoref{fig:query-rewrite} shows examples of rule specification recipe and \QA recipe. Note that the rewrite \QA may elaborate on application of multiple rules, and thus \QA recipes can assist in guiding rule selection by considering the interrelations among these rules. Besides, rule specification recipes, derived from various sources like documents and codes, provide complementary insights to support query rewrite, as illustrated in \autoref{tab:rewrite-rule-spec}.

In this paper, we focus on how to prepare the \QA repository and rewrite rule specification repository (see Section~\ref{sec:evidence-prepare}), how to generate \QA recipe and rule specification recipe (see Section \ref{sec:evidence-retrieval}), and how to use them to rewrite a query (see Section \ref{sec:llm-rewrite}).

\vspace{-.75em}
\subsection{Large Language Models}
\label{sec:large-language-models}
\vspace{-.25em}

Since LLMs have hallucinations~\cite{huang2023survey}, retrieval-augmented generation (RAG) has been proposed to mitigate this issue by indexing task-specific knowledge, retrieving relevant content for a given query, and generating answers based on the retrieved context~\cite{gao2023retrieval,zhang2025sage,zhao2024chat2data,li2024llm}.
However, existing RAG techniques face limitations when applied directly to query rewrite for two primary reasons. First, there is absence of embedding methods capable of accurately evaluating the similarity between \QAs and the input query. For example, the conventional RAG technique relies on text embeddings for both the \QA and the query, capturing only their semantic similarity while neglecting structural information pertinent to query rewrite. To address this gap, we propose a hybrid structure-semantics approach for retrieving relevant \QAs, as detailed in Section \ref{sec:evidence-retrieval}.
Second, the multitude of rules presents a significant challenge for \llms, as directly arranging these rules can lead to serious hallucination problems. This necessitates the development of a task-specific, step-by-step \llm algorithm that breaks down the query rewrite process into simpler, more manageable stages, as elaborated in Section \ref{sec:llm-rewrite}.

\label{sec:overview}
\begin{figure*}
	\centering\vspace{-2em}
\includegraphics[width=0.95\linewidth]{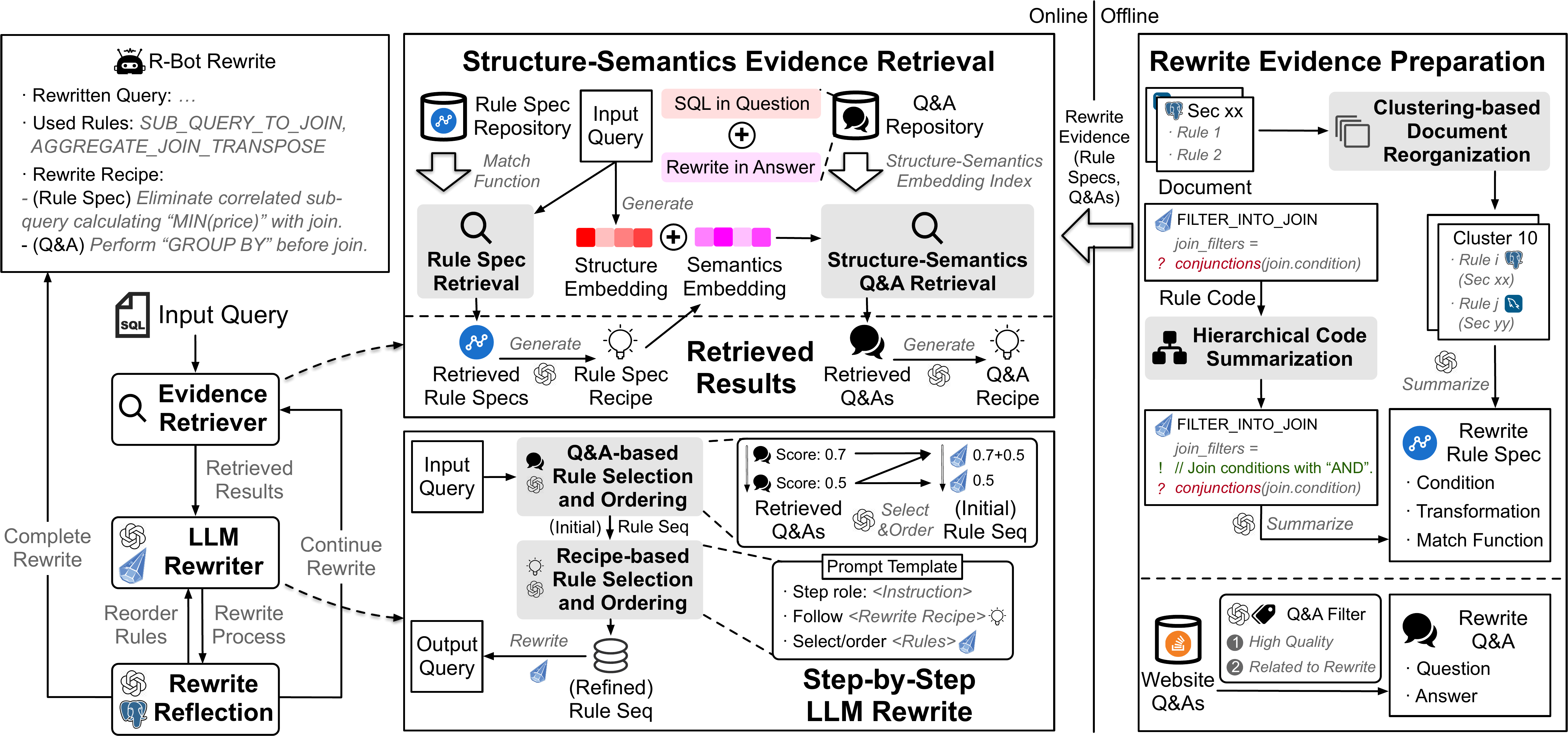}
    \vspace{-1.25em}
	\caption{Overview of \oursys.}
	\label{fig:overview} \vspace{-1.75em}
\end{figure*}

\vspace{-.5em}
\section{The Overview of \oursys} \label{sec:overview}
\vspace{-.25em}

\oursys includes an offline stage and an online stage (see \autoref{fig:overview}).

\hi{Offline Rewrite Evidence Preparation.} This stage aims to extract rewrite \QAs from the Web and rewrite rule specifications from rewrite codes and database documents, subsequently storing them in \QA repository and rule specification repository, respectively. 
Rewrite rule specifications are general and apply to multiple SQL queries, while rewrite \QAs are specific and pertain to individual queries. 
These resources assist in rewriting an online SQL query {by guiding \llm-based selection of rewrite rules, without compromising the SQL equivalence inherently ensured by the rules.} 

We extract rewrite rule specifications from two types of resources.  $(i)$ Database menus and documents. Given that query rewrite evidence is often dispersed across various documents, sections, and paragraphs, it becomes necessary to aggregate these evidences from diverse sources through semantic clustering and distill them into a coherent rewrite rule;  $(ii)$ Rule codes. Considering the complexity of the code, characterized by its intricate nested calls, we employ a hierarchical strategy to streamline the code from simple to complex. This approach involves initially summarizing straightforward functions, followed by recursively summarizing more complex functions. In this process, symbol references are clarified using functions that have already been summarized. Ultimately, this method results in a concise summary of the overall rule code, effectively condensing the rule into a simplified format.

We extract \QAs from  website \QAs (e.g., Stack Overflow~\cite{stackoverflow}).  Given the wide range of topics and variable quality of these website \QAs, we filter out high-quality \QAs related to query rewrite, by SQL selections (e.g., on question tags) and \llm filtering. 

Besides, we construct efficient indexes to enhance the performance of \QA and rule specification retrieval. We will discuss the detailed techniques in Section~\ref{sec:evidence-prepare}.

\hi{Online LLM-Guided Query Rewrite.} Given an online SQL query, this stage retrieves pertinent \QAs and rule specifications, leverages them to rewrite the query, and offers reflections that not only complete the rewrite process but facilitate its further refinement.
While \llm may have encountered data resembling rewrite evidences during pre-training, their sparse density often leads to hallucination in query rewrite.
To address this, we guide \llm rewrite using the pertinent rewrite evidences prepared in a concise format.

\emph{\textbf{(1) Structure-Semantics Evidence Retrieval.}} For an online SQL query, we retrieve relevant evidences from rewrite rule specification repository and \QA repository. 

$(i)$ Rule specification retrieval. As the rule specification has matching conditions and matching functions, we can easily retrieve the relevant rule specifications whose matching functions are satisfied by the input SQL query;

$(ii)$  \QA retrieval. There are two types of \QAs potentially relevant to the SQL query. First, the SQL questioned in \QA structurally matches the input SQL query. We require a structure-aware matching method to retrieve such \QAs. To this end, we propose a query structure embedding composed of $(i)$ the query template embedded by pre-trained embedding, and $(ii)$ one-hot embedding which represents the query's matched rule specifications to assess the structural similarities. Second, \QA semantically matches the input SQL query, such as the rewrite explanations provided in the answer section of the \QA. However, SQL usually has no natural language rewrite explanations. To this end, we can leverage the retrieved rule specification to identify relevant \QAs based on their semantics. We propose a semantics embedding method to embed the rule specification and \QA, and then assess their similarities based on a similarity function (e.g., $L^2$-distance~\cite{chroma}).  Lastly, to retrieve \QAs relevant to the input query both structurally and semantically, we merge the query’s structural embedding with its semantic embedding into a unified representation. We then build an embedding index for \QA repository offline. For an online query, we utilize this index to efficiently retrieve top-$k$ relevant \QAs.

For retrieved rule specification and \QA of the input SQL query, we leverage \llm to generate SQL-aware rule specification recipe and \QA recipe that describe how to utilize them to rewrite this SQL query.  We will discuss the technical details in Section~\ref{sec:evidence-retrieval}.  

The retrieved rule specification recipe and \QA recipe will be used to reformulate the SQL query in the following steps. 

\emph{\textbf{(2) Step-by-Step \llm Rewrite.}} Given the rewrite rules supported by the query rewrite engine (e.g., Apache Calcite~\cite{calcite}), we direct \llm to select pertinent rules to rewrite. As there are many rules, if we directly instruct \llm to arrange a rule sequence, \llm can encounter serious hallucination problem. To mitigate this problem, we propose a step-by-step \llm rewrite method that decomposes rule-based query rewrite into several simpler steps.

$(i)$ \QA-based rule selection and ordering. Given the rewrite rules and a sorted sequence of \QAs retrieved from the previous step, ranked by the relevance score, we select and rank the rules. We first initialize the score of each rule as 0. Then for each pair of a rule and a \QA, we use \llm to evaluate whether they are relevant, i.e., whether the rule is applicable in light of the \QA. If applicable, we increase the score of the rule by the score of this \QA. Then by enumerating all the pairs of rules and \QAs, we can get the final score of each rule and rank the rules based on the final score. Since the relevance can be evaluated by \llm offline, this step can be efficiently executed by algorithms. 
    
$(ii)$ Recipe-based rule selection and ordering.
Building on the preliminary rule sequence established in the previous step, we utilize \llm to sift through and exclude any rules that do not align with the recipe. A straightforward way is to enumerate each pair of recipes returned by step (1) and rules returned by step $(2.i)$, and ask \llm to evaluate their relevance and rank the rules. However, there are two limitations. First, it may overlook the rule relevance, since a recipe may encompass multiple rules. Second, since the recipe is generated from queries and cannot be evaluated offline, assessing each pair with \llm becomes costly. To address this issue, we propose a filtering method to efficiently select the rules. 
    
$(iii)$  Rule-based rewrite. Based on the selected rules, we input them to the query rewrite engine to rewrite the query.

We will discuss the technical details in Section~\ref{sec:llm-rewrite}.

\emph{\textbf{(3) Rewrite Reflection.}} It provides rewrite reflections to either further refine the query or to finalize the rewrite process. It has two reflection resources. The first involves getting the cost of the rewritten query from databases, comparing it with the cost of the query prior to its rewrite, and returning {\it complete} if the cost of the rewritten query is smaller; {\it continue} otherwise.~\footnote{{Database statistics play a crucial role in query optimization. 
However, \llm often struggles to directly understand complex database structures and intricate statistical data. 
To address this, we leverage the statistics indirectly through query costs, utilizing these costs to guide the \llm reflection mechanism.}} The second involves asking \llms to check whether or not all the rewrite recipes are realized by the query rewrite in this step, and returning {\it complete} if yes; {\it continue} otherwise. So there are four possible reflections. 
    
$(i)$     {\it complete, complete:} It finalizes the rewrite process and returns the rewritten query.
  
$(ii)$ {\it complete, continue:} It further rewrites the query by jumping to step (1) with the previously rewritten query as the input. Specifically, it starts a new round of \llm-guided query rewrite, where new rewrite evidences can be retrieved and new rules can be selected based on the previously rewritten query. 
    
$(iii)$ {\it continue, complete:} It further refines the query by jumping to step $(2.ii)$, focusing on reordering the existing set of rules. Specifically, besides the recipes and the rules, we further input the rules actually used in the previous rewrite process, and instruct \llm to prioritize the unused rules.
    
$(iv)$ {\it continue, continue:} This approach integrates elements from branches $(3.ii)$ and $(3.iii)$. Initially, it defaults to branch $(3.iii)$ unless this path is revisited excessively, surpassing a predefined threshold. Under these circumstances, given that branch $(3.iii)$ has exhaustively explored the rules for the current query, the process transitions to branch $(3.ii)$. This shift initiates a new cycle, aiming to further refine the query.

\hi{Deployment at Huawei.}
We have deployed \oursys at Huawei database GaussDB~\cite{li2021opengauss,li2024gaussdb}.
Initially, GaussDB identifies slow SQL queries, such as those with execution time exceeding one minute. Then, GaussDB utilizes \oursys to rewrite the detected query. Next, if the version rewritten by \oursys proves to be more efficient than the one rewritten by GaussDB, the system caches the query pattern using its domain-specific language (DSL). This pattern is integrated into GaussDB non-intrusively via a SQL-like plugin, which activates instantly without requiring database version updates or code changes, ensuring a seamless experience for front-end applications. During runtime, if a new query matches the cached pattern, GaussDB transparently invokes \oursys to rewrite the query and apply the optimization. Furthermore, \oursys has been widely used to address slow SQLs at Huawei. We have also validated \oursys on a real-world dataset from China’s largest bank (ICBC) (see Section \ref{sec:deploy}). The results demonstrate that \oursys effectively optimizes real queries, significantly improving latency of 14 critical slow queries and reducing overall latency from 9.23 hours to 4.37 hours. 

\vspace{-.5em}
\section{Rewrite Evidence Preparation}
\label{sec:evidence-prepare}
\vspace{-.25em}

We discuss how to extract and standardize rewrite evidences from diverse rewrite sources. This evidence is crucial for crafting a comprehensive rewrite recipe that guides the \llm rewrite process. We explain respectively how to prepare rewrite rule specifications (see Section \ref{sec:rule-prepare}) and rewrite \QAs (see Section \ref{sec:qa-prepare}).

\vspace{-.75em}
\subsection{Rewrite Rule Specification Preparation}
\label{sec:rule-prepare}
\vspace{-.25em}

The rewrite rule specification clearly outlines, in natural language, the condition for use, the query transformation operations to be executed, and the matching function used for evaluation. It contains three key components. 
$(i)$ ``condition'': a prerequisite that a query must fulfill to utilize the rule; 
$(ii)$ ``transformation'': detailed steps for transforming the query into an equivalent form that is optimized for more efficient execution;
and $(iii)$ ``matching function'': an executable function that outputs `1' if the input query matches the rule; and `0' otherwise.

Generating rewrite rule specifications is challenging due to the considerable effort needed to distill and synthesize information from various sources into a concise format. This process includes summarizing extensive rewrite codes, which may span thousands of lines and feature complex structures, as discussed in Section \ref{sec:code-prepare}. Additionally, it integrates crucial but scattered information from documents on rewrite rules, as outlined in Section \ref{sec:doc-prepare}.

\vspace{-.5em}
\subsubsection{Transforming Rule Code into Rule Specification}
\label{sec:code-prepare}

\begin{figure}[!t]
	\centering
\includegraphics[width=.9\linewidth]{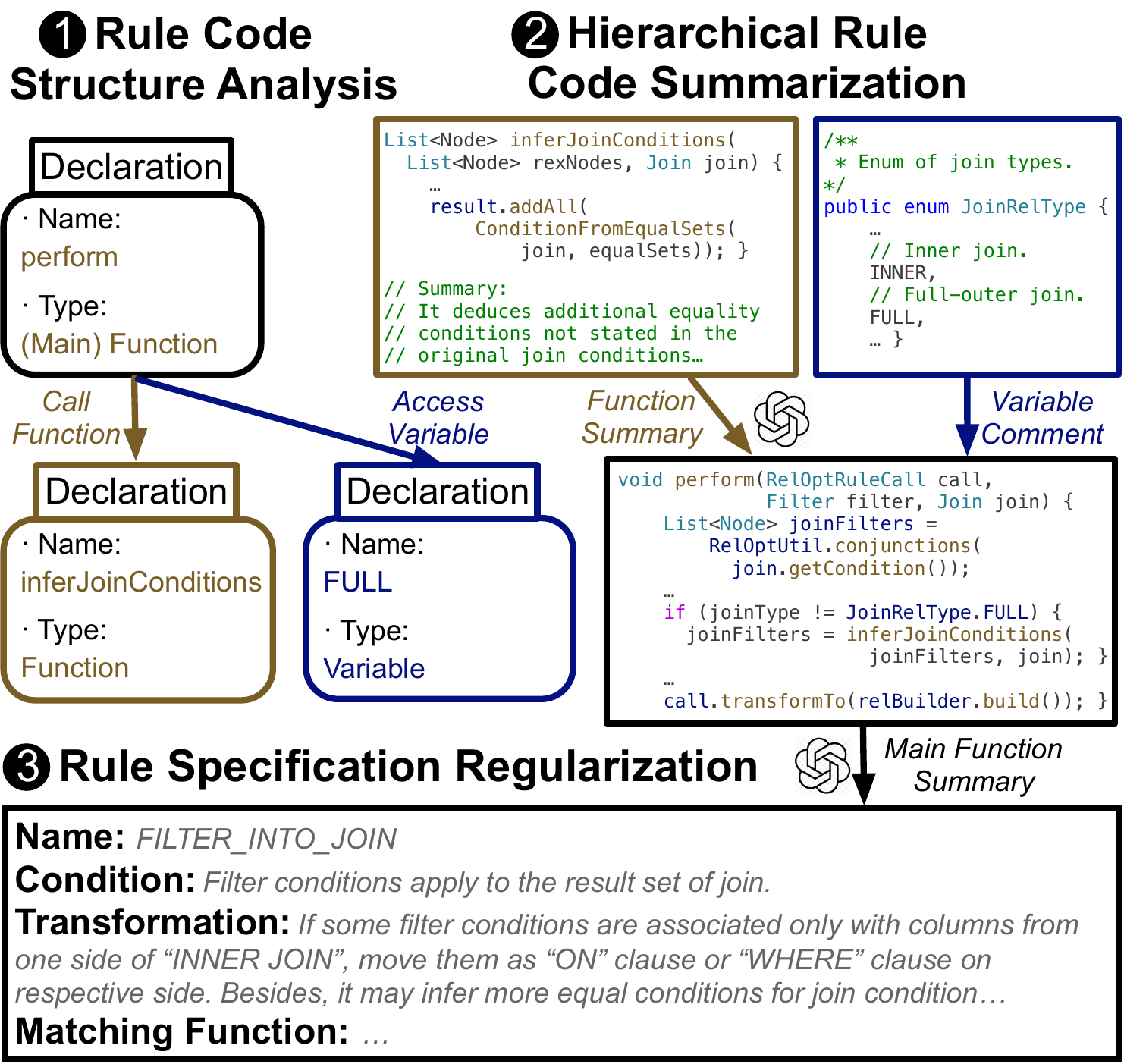}
 \vspace{-1em}
	\caption{Transforming Rule Code into Rule Specification.}
	\label{fig:code-prepare}
 \vspace{-1.75em}
\end{figure}

Given that some query rewrite engines, such as Apache Calcite~\cite{calcite}, are not accompanied by comprehensive documentation, we are compelled to decipher the rewrite rules directly from the raw code. This process involves navigating through complex code structures that include intricate nested calls. To address this challenge, we introduce a hierarchical rule code summarization method, as illustrated in \autoref{fig:code-prepare}. Our approach begins with the construction of a rule code structure tree, emanating from the rule's main function. In this tree, each node represents a symbol declaration (e.g., functions, variables, classes), while each edge denotes a symbol reference relationship. Progressing through the structure, we methodically summarize the declaration code, moving from simple to more complex elements and clarifying symbol references using previously summarized symbols. With the summary of the root node at our disposal, we guide \llm to convert this into a standardized rewrite rule specification.

\hi{Rule Code Structure Analysis.}
Given the symbol references of the rule code, it is crucial to clarify the symbol declarations before summarizing the rule code. We first build a root node representing the main function. Then, we use code analysis tools (e.g., JavaSymbolSolver~\cite{javasymbolsolver}) to associate the symbols with their corresponding declarations, which are children nodes of the root node. If the declaration of some nodes also accesses other unseen symbols, we further resolve its symbol references. We recursively expand the nodes until reaching  built-in symbols. In this way, we obtain a rule code structure tree, where each node represents a symbol declaration and each edge represents a symbol reference relationship.

\hi{Hierarchical Rule Code Summarization.} 
The complex rule code structure poses two challenges for \llm summarization.
First, due to the relevance of nearly every declaration in the code, if we directly input them to \llm, the long context can greatly degrade \llm performance~\cite{liu2024lost,li2023long}. 
Second, the substantial width and depth of the code structure (e.g., tens of nodes) further increase the reasoning burden. 
To address these issues, we propose a hierarchical rule code summarization method. First, if a node declaration already has detailed comments, summarization is unnecessary. Second, leaf node without comments can be directly summarized by \llm due to the simple code and absence of unfamiliar symbols. 
Third, for the non-leaf node whose children are already summarized, the symbol references in the declaration code can be clarified by its children summaries. Specifically, we insert the symbol summaries as comments into the declaration code, which enables accurate \llm summarization. This process is repeated recursively until the root node is summarized, yielding a summary of the entire rule code.

\hi{Rule Specification Regularization.} 
To facilitate \llm understanding, we regularize the rule code as a standard rewrite rule specification. Specifically, we use \llm to extract the condition and transformation of the rule, using the prompt \textit{$p_{reg}=$``Given a rewrite rule code summary, your task is to extract the rewrite rule that  explains completely and detailedly the condition and transformation.''}.

\vspace{-1em}
\subsubsection{Transforming Rewrite Document into Rule Specification}
\label{sec:doc-prepare}

Considering the variety of rewrite documents, such as those for PostgreSQL and MySQL, note that sections within a single document may cover rewrite rules that bear weak relation to one another. For example, optimizations for the ``WHERE'' clause might discuss both constant folding and index utilization without clear interrelation. Additionally, components complementary to a rule can be dispersed across different documents. For instance, while a MySQL document might detail conditions conducive to acceleration via index utilization, a separate PostgreSQL document could highlight how certain column transformations might inhibit the use of indexes. Together, these insights from disparate sources can contribute to forming a comprehensive rewrite rule specification. 

To address this, we propose a clustering-based document re-organization method. First, we use \llm to extract rewrite rules from rewrite documents. Second, for extracted rules, we cluster the correlated ones together into one group, where we evaluate their semantics similarities by their text embeddings (e.g., SBERT~\cite{reimers2019sentence}). Third, we use \llm to summarize each rule cluster, and transform each cluster summary as a regularized rewrite rule specification. 

\hi{Rule Extraction.}
We use \llm to extract rewrite rules from the rewrite documents in two steps. First, if we directly input all the documents to \llm, it often overlooks important details in the middle of the extremely long context (e.g., 100k)~\cite{liu2024lost,li2023long}. We thus split the documents into structured blocks (e.g., sections, sub-sections) that each can be effectively processed by \llm. Second, we instruct \llm to extract rewrite rules from the split blocks. To mitigate the hallucination problem~\cite{huang2023survey}, we require \llm to locate supporting content in the source document. If no such content can be found, we can deem the extraction low quality and repeat \llm extraction.

\hi{Rule Clustering.}  To identify pertinent rules among documents (e.g., condition push-down involved in both PostgreSQL and MySQL documents), we cluster the rules based on rule semantics.
Specifically, we embed each rule into vectors (e.g., using multi-qa-mpnet-base-cos-v1~\cite{reimers2019sentence,multiqampnet}), and cluster them using Gaussian Mixture Models~\cite{sarthiraptor}. To enhance clustering in high-dimensional space, we apply Uniform Manifold Approximation and Projection for dimensionality reduction by approximating local data manifold~\cite{mcinnes2018umap}.

\hi{Rule Specification Regularization.}
We use \llm to obtain standardized rule specifications from rule clusters. First, we summarize each cluster using \llm  with prompt \textit{$p_{summ}=$``Given rewrite rule components, your task is to summarize them into one paragraph, and your summary should include as many details as possible.''}  
Then, we use \llm to transform the summary to regularized rewrite rule specification, by extracting rule condition and transformation.

\vspace{-.75em}
\subsection{Rewrite \QA Preparation}
\label{sec:qa-prepare}
\vspace{-.25em}

The rewrite \QA showcases how to rewrite a particular SQL query to an optimized one, which is composed of two parts: 
$(i)$ ``question'' that denotes a query rewrite request; 
$(ii)$ ``answer'' that provides the query transformations for the question.
There are plenty of rewrite \QAs within online database community forums (e.g., millions at Stack Overflow~\cite{stackoverflow}), often covering practical cases beyond standard rewrite rules. 
To filter high-quality rewrite \QAs from these mixed sources, we use a hybrid method: first selecting by question tags (e.g., ``query-optimization'') and community feedback (e.g., Stack Overflow score higher than 3), and then using \llm to verify their relevance to query rewrite.

\vspace{-.5em}
\section{Structure-Semantics Retrieval}
\label{sec:evidence-retrieval}
\vspace{-.25em}
Considering the existence of vast repositories of rule specifications and \QAs, only a minimal fraction of these resources is pertinent to an online SQL query. Thus, there is a need to identify the relevant ones both effectively and efficiently.  In this section, we introduce how to retrieve relevant rewrite evidences (including rule specifications and \QAs). First, we retrieve relevant  rule specifications using a function-based rule retrieval method (see Section ~\ref{sec:rule-retrieval}).  Second, we retrieve relevant rewrite \QAs using a hybrid structure-semantics method, with both query structures and rewrite semantics aligned with the input query (see Section ~\ref{sec:qa-retrieval}).  Lastly, we generate tailored rewrite recipes for the input query by leveraging both the retrieved rule specifications and \QAs (see Section ~\ref{sec:qa-recipe}).

\vspace{-.75em}
\subsection{Rewrite Rule Specification Retrieval}
\label{sec:rule-retrieval}
\vspace{-.25em}

Given an input SQL query $q$, we examine each rule specification and apply its associated matching function to get a boolean value which indicates whether the condition of the rule is satisfied. If the result is true, the corresponding rule specification $rs$ is retrieved. Then, we use \llm to generate a rule specification recipe,  using a prompt \textit{$p_{rule\_spec\_recipe}=$``Given an SQL query `$q$' and a rewrite rule specification `$rs$', your task is to explain concisely and detailedly how the rule applies to the query, by specifying (1) the SQL segments matched by the condition, and (2) the transformation of the rule.''}

\begin{figure}[!t]
	\centering
\includegraphics[width=.95\linewidth]{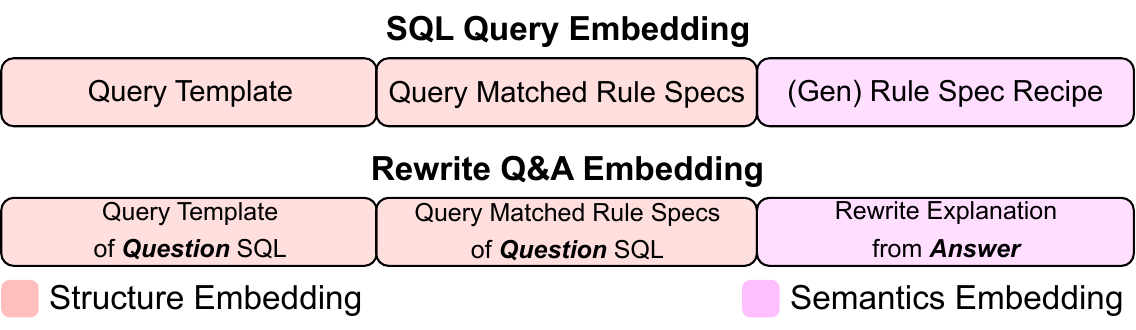}
        \vspace{-1.5em}
	\caption{Structure-Semantics Embeddings in \QA Retrieval.}
	\label{fig:qa-retrieval}
 \vspace{-1.75em}
\end{figure}

\vspace{-.75em}
\subsection{Rewrite \QA Retrieval}
\label{sec:qa-retrieval}
\vspace{-.25em}

There are two types of \QAs that could potentially enhance the query rewrite. Firstly, the questions within \QAs exhibit a high structural similarity to the input query, indicating that the answers of \QAs have the potential to assist in rewriting the query. Second, the answers within \QAs demonstrate a high semantic similarity to the input query. This suggests that the \QAs are capable of addressing similar issues or bottlenecks present in the input query, such as those involving a sub-query with an aggregate function.

\hi{Structure-Semantics Embeddings.}  To effectively identify these two types of \QAs, we propose structure-semantics embeddings for both SQL queries and \QAs (see \autoref{fig:qa-retrieval}). (1) \textit{SQL Query Embedding.} First, we introduce a structure-aware query embedding strategy, including $(i)$ generating query templates embedded by pre-trained embedding (e.g., text-embedding-3-small~\cite{openai}), and $(ii)$ generating a one-hot embedding to indicate which rule specifications match the SQL query. The two embeddings capture the essential structural features to query rewrite (see Section \ref{sec:query-structure-embed}). Second, we propose a semantic matching method designed to discern the semantic similarity between the SQL query and \QAs. However SQL may not encapsulate sufficient semantic information. Fortunately, the retrieved rule specification recipe of SQL in Section \ref{sec:rule-retrieval} encompasses these semantics, enabling us to derive an embedding from it. Third, we concatenate the structural and semantic embeddings to obtain a combined structure-semantics embedding. Given the possibility of retrieving multiple rule specification recipes, each SQL query has the potential to generate multiple embeddings.
(2) \textit{Rewrite \QA Embedding.} Similarly, we generate embeddings of \QAs. For each \QA, its embedding includes a structure-aware embedding of its query in the question part and a semantic embedding of its answer part. These embeddings allow us to identify \QAs that exhibit a high similarity to the input query in terms of their embeddings.

\hi{Structure-Semantics \QA Retrieval.}  To optimize performance for structure-semantics \QA retrieval, a unified structure-semantics embedding index (e.g., HNWS~\cite{chroma}) is constructed for \QAs offline. Given an SQL query, we generate its structure-semantics embedding and retrieve relevant \QAs using the index (see Section \ref{sec:qa-retrieval-details}).

\vspace{-1em}
\subsubsection{Structure-Aware SQL Query Embedding}
\label{sec:query-structure-embed}
Existing query embedding models~\cite{tang2022preqr,zhao2022queryformer,zhao2023comparative,openai} cannot be directly applied for embedding query structure for query rewrite due to two key limitations. On one hand, the query-specific embeddings mostly adopt small-scale neural networks (e.g., lower than 0.1B) fine-tuned on narrow datasets~\cite{zhao2022queryformer,zhao2023comparative} or workloads~\cite{tang2022preqr}, limiting their generalizability.
On the other hand, the general text embeddings (e.g., text-embedding-3-small~\cite{openai}) fail to effectively represent query structure, as they are sensitive to irrelevant information:
$(i)$ The identifiers (e.g., table/column names) and literals in the query often disrupt characterization of query structure. For example, while renaming the schema in a query can maintain the integrity of the query rewrites, this action results in a version with significantly altered semantics and it generates an embedding distinctly different from that of the original query; $(ii)$ Since query rewrites typically focus on transforming only certain parts of the query (e.g., a sub-query), disregarding other irrelevant parts can help clarify the query structure; $(iii)$ The text embedding models cannot guarantee commutative invariance, implying they may generate dissimilar embeddings for expressions such as ``$p_1\ AND\ p_2$'' and ``$p_2\ AND\ p_1$''. These can be addressed through carefully designed textual transformations.

To address the aforementioned challenges, we propose a structure-aware query embedding method, where each SQL query embedding is composed of two parts. The first part distills the core query structure as query templates, and uses pre-trained text embedding to embed them. The second part employs a one-hot encoding strategy to indicate whether the query matches a rule specification, with `1' representing a match and `0' indicating no match. The width of this embedding corresponds to the total number of rule specifications. When comparing the two query structure embeddings, the first one captures a more global representation of the query structure, whereas the second focuses on more local expressions.

\hi{Dataflow Based Query Template Embedding.} 
To capture the essential SQL features for query rewrite, we initially reformulate the SQL query into composite dataflows that detail operations on the involved identifiers. Specifically, given a query $q$, a dataflow $\delta$ is a list of SQL operations sequentially performed on one or multiple tables or columns in the query, which corresponds to an SQL segment of the query, e.g., logical expression, mathematical expression, ``WHERE'' clause, ``GROUP BY'' clause, ``FROM'' clause, ``JOIN'' clause, ``SELECT'' list, sub-query, etc. For instance, we show sample dataflows for the input query in \autoref{fig:query-rewrite}, concentrating on the column ``$bonus.sal$'': 
$
\{sal, SELECT\ MAX(sal), emp.sal = ANY\ (SELECT\ MAX(sal)\ ...)\}
$
where ``$sal$'' is first input from ``$bonus$'' table, projected with an aggregate function ``$MAX(\cdot)$'', and then compared with the column ``$emp.sal$'' from the outer query.

Since we focus on logical query rewrite, dataflows can be considered independent if they do not involve common identifiers.  Then, if we examine the rewrite rule from dataflow perspective, the condition must restrict the pattern of certain dataflows associated with a specific identifier. Otherwise, there are scarcely any potential query rewrites within a set of independent dataflows.  Consequently, we can generate query templates that concentrate on the SQL operations related to each particular identifier.

To ensure query templates are robust to schema renaming, we propose an isolated masking method. If a query template involves multiple identifiers, swapping any two identifiers alters its semantics and disrupts the embedding. Thus, we derive multiple query templates from the query, with each template preserving only one identifier while masking the others. We then refine the templates in three steps. First, to limit the number of templates derived from complex queries, we retain only those where the SQL operations associated with the identifier are likely to match certain rewrite rules. Second, we simplify the query template by eliminating SQL operations that do not involve the key identifier. Third, for the commutative SQL operands, we sort them lexically, thereby rendering the template invariant to operand order.

{\it Step 1: Potential Identifier Selection.}
Considering the SQL operations associated with the identifier, we observe that multiple appearances of the same identifier in mutually exclusive dataflows may prompt query rewrites by leveraging operation correlations. For instance, the twice appearance of column ``$a$'' in ``$a<b$'' and ``$a=5$'' can indicate constant folding, replacing ``$a<b$'' with ``$5<b$''.  Then, given an SQL query, we propose two ways to measure identifier frequency: $(i)$ Column appearances: By traversing the query, we count how often each column appears. Note that columns appearing as direct projections (e.g., ``SELECT a") are not counted, as this does not involve any meaningful SQL operations. 
$(ii)$ Table appearances: We count the appearances of the table which is utilized either in the ``FROM'' clause or within a ``JOIN'' operation. 
With the frequency of tables and columns, we focus on identifiers that appear more than once within the query. We then select their corresponding query templates as representative, ensuring we capture the most significant patterns for query rewrite.

{\it Step 2: Query Template Reduction.}
Given the single identifier preserved in the query template, we standardize it as ``$table$'' for table and ``$column$'' for column. Moreover, to remove irrelevant query information, we mask other identifiers with ``\_'' and the literals with ``?''. We make a distinction between identifier mask and literal mask, since they indicate different query rewrite potentials. For instance, we can preserve the constant folding pattern in the query template with ``$column<\_\ AND\ column=?$'', which means the constant equality can be transferred to other conditions involving ``$column$''. Then, we further simplify the query template following three steps.  
$(i)$ If any dataflow in the query template involves only masked literals ``?'', no identifiers, and no non-deterministic functions, the dataflow can be regarded as a literal and replaced with literal mask ``?''.
$(ii)$ If any dataflow involves some identifiers but no explicit identifiers, we replace it with identifier mask ``\_'', as it is independent to the potential identifier. 
$(iii)$ For the dataflow of clause like ``$t_1$ JOIN $t_2$ ON $c$'', if  the join condition ``$c$'' and a join table (e.g., ``$t_2$'') are both masked with ``\_'', we replace the clause with ``$t_1$'' to remove irrelevant details.
These steps are repeated to streamline the query template until it cannot be further matched, leaving a structure that retains only the essential core related to the potential identifier.

{\it Step 3: Commutative Invariance Guarantee.}
We further transform the query template for commutative invariance. Specifically, we search for commutative operators in the query template (e.g., addition, set intersection), and sort their operands lexically (e.g., ``$column<\_$'' before ``$column=?$''). Equipped with the refined query templates, we then obtain query template embeddings with pre-trained text embedding (e.g., text-embedding-3-small~\cite{openai}).

\hi{One-Hot Embedding for Matched Rule Specifications.}
Given an SQL query, its matched rule specifications also reflect query structure. For instance, queries matching ``SUB\_QUERY\_TO\_JOIN'' rule likely exhibit similar structures around sub-queries. As shown in Section \ref{sec:rule-retrieval}, we first examine the rule specifications with their associated matching functions. Following this, we construct a one-hot encoding from the rule matching results, where each position corresponds to a rule specification. If a rule specification matches the query, its corresponding position is `1'; `0' otherwise. The one-hot embedding for matched rule specifications can capture local structural information indicated by the rule condition, complementary to the global structures captured by query template embedding.

\vspace{-.5em}

\subsubsection{Structure-Semantics \QA Retrieval}
\label{sec:qa-retrieval-details}
First, we build a structure-semantics embedding index for \QA repository in three steps.
$(i)$ We extract the queries in the \QAs, generate query templates, and embed them with pre-trained text embedding (e.g., text-embedding-3-small~\cite{openai}). We also match the queries with rule specifications to build one-hot embedding for matched rule specifications. $(ii)$ We refine the \QA by preserving the semantics of query rewrite and eliminating irrelevant text with the help of \llm, then embedding the streamlined information using a pre-trained text embedding model (e.g., text-embedding-3-small~\cite{openai}). $(iii)$ We normalize the three embeddings as unit vectors,  concatenate them to form a holistic embedding, and insert them into the index. Next, for an online SQL query, we similarly generate its embedding and identify the most relevant \QAs with top-$k$ embedding similarities.

Every SQL query may be associated with multiple templates, leading to multiple structural embeddings. Similarly, each SQL query encompasses multiple specification recipes, resulting in multiple semantic embeddings. We concatenate these embeddings for all possible combinations and, for each concatenated embedding, we utilize the index to identify top-$k$ \QAs with the highest similarities.  To combine the retrieved results from multiple embeddings, we adopt the Reciprocal Rank Fusion (RRF) method~\cite{cormack2009reciprocal}. Specifically, for each retrieved list corresponding to an embedding, we assign a score to the $i$-th \QA in the ranked list as, $s(qa_i)=\frac{1}{\alpha+i}$, where $\alpha$ is 60 by default. The RRF score of a \QA across all the retrieved lists is calculated by summing its scores from each list (assigning a score of 0 if the \QA is absent in a list). We then identify the \QAs with highest top-$k$ RRF scores as the final selection, which comprehensively captures the structure and semantics relevance.

\vspace{-.5em}
\subsection{Rewrite Recipe Generation}
\label{sec:qa-recipe}
\vspace{-.25em}

For an input SQL query $q$, since the retrieved \QA $qa$ is used to rewrite similar but different queries, we first use \llm to generate rewrite recipes describing how to rewrite the input query inspired by the \QA, using the prompt \textit{$p_{qa\_recipe}=$``Given an SQL query  `$q$' and a rewrite Q\&A `$qa$', your task is to propose some strategies on rewriting the query, by (1) transferring the \QA strategy to the query, and (2) explaining the strategy detailedly.''}  Second, to integrate both rule specification recipes (see Section \ref{sec:rule-retrieval}) and \QA recipes, we condense those that are closely related and eliminate any duplicates.
Thus, we utilize \llm to cluster them by their semantic similarity, and summarize each recipe cluster concisely into a single recipe.

\vspace{-.5em}
\section{Step-by-Step \llm Rewrite}
\label{sec:llm-rewrite}
\vspace{-.25em}

Given a rule-based query rewrite engine (e.g., Apache Calcite~\cite{calcite}), we utilize the retrieved \QAs and derived rewrite recipes to assist \llm in selecting appropriate rewrite rules from the engine to rewrite the input SQL query. Given that the number of possible rule sequences grows exponentially with the number of rules, \llms are susceptible to errors when choosing from a vast array of rules. To address this challenge, we introduce a step-by-step filter-refinement method designed to meticulously select high-quality rules. Specifically, we start with a filtering step by employing an efficient \QA-based method for rule selection and ordering, allowing us to preliminarily arrange the rules. Next, we instruct \llm to select the rules and arrange the order according to the recipes. Then, we feed the selected rules into the query rewrite engine to rewrite the query with the rules. Lastly, we evaluate the outcomes of the rewrite process to determine whether further refinement is needed or if the query rewrite can be considered complete. 

\hi{Step 1: \QA-based Rule Selection and Ordering.}
Given the rewrite rules and the input query, we filter the rules and arrange the order in three steps.
First, we select the rules directly matched by the query, and assign them a relevance score according to their transformation to the query. Second, we select the rules indirectly relevant to the query using the retrieved \QAs. Based on the retrieval scores of the \QAs, we assign each indirectly relevant rule with a relevance score. Third, we rank the rules by their relevance scores in a descendent order, which serves as an initial order. 

$(i)$ We first filter rules whose matching functions are satisfied by the input query. However, the matched rules vary in relevance based on their actual transformation. Specifically, the rules of the query rewrite engine can be classified into two types of transformations: the normalization rule (e.g., ``SUB\_QUERY\_TO\_JOIN''), which nearly always reduces query cost; and the exploration rule (e.g., ``AGGREGATE\_JOIN\_TRANSPOSE''), which transforms the query but does not consistently result in cost reduction~\cite{taft2020cockroachdb}. We classify the rules based on expert experience, and assign relevance scores to the matched rules based on their transformation types. Initially, every rule has a score of $-\infty$. Then, the matched normalization rules are assigned $+\infty$ as closely relevant, and the matched exploration rules are assigned $0$ as weakly relevant. 

$(ii)$ Besides the directly relevant rules, we also detect indirectly relevant rules using the retrieved \QAs. Specifically, for each pair of rule and \QA, we leverage \llm to evaluate whether the rule can be applied in the context of the \QA. If applicable, the rule's score is incremented by the \QA similarity score to the input query. Note that if the rule has a score of $-\infty$, we first initialize it as $0$ before increment. Enumerating all the pairs of rules and the retrieved \QAs, we obtain the final score of each rule $r$. {\it Note that the relevance can be performed by \llm offline for each possible pair of rule and \QA, thus not affecting the algorithm’s efficiency.}

$(iii)$ Given the rules with relevance scores, we first filter out those with $-\infty$ as irrelevant. Then, we order the remaining rules by prioritizing the rules with higher scores, which comprehensively reflect their direct and indirect relevance to the input query. 

\begin{table*}[!t]
\centering\vspace{-2em}
\caption{Comparison of Query Latency.}\vspace{-1em}
\label{tab:methods-execution-time}
{
\scriptsize 
\begin{tabular}{cccccccccc}
\hline
\multirow{2}{*}{Query Latency (s)} & \multicolumn{3}{c}{TPC-H 10x}                                                                                                                                                              & \multicolumn{3}{c}{DSB 10x}                                                                                                                                                                                  & \multicolumn{3}{c}{Calcite (uni)}                                                                                                                                                                              \\ \cline{2-10} 
                                    & Average                                                              & Median                                                     & p90                                                      & Average                                                              & Median                                                              & p90                                                               & Average                                                              & Median                                                              & p90                                                               \\ \hline
Origin                              & 104.86                                                            & 10.60                                                   & 300.00                                                   & 37.76                                                             & 5.28                                                             & 300.00                                                            & 109.73                                                             & 56.35                                                            & 300.00                                                             \\
\hline
\lr & 69.60 ($\downarrow$33.6\%) & 12.26 ($\uparrow$15.7\%) & 300.00 ($\downarrow$0.0\%) & 30.47 ($\downarrow$19.3\%) & 5.28 ($\downarrow$0.0\%) & 55.02 ($\downarrow$81.7\%) & 79.07 ($\downarrow$27.9\%) & 5.24 ($\downarrow$90.7\%) & 300.00 ($\downarrow$0.0\%) \\
\vanillathree                  & 85.98 ($\downarrow$18.0\%)          & 10.60 ($\downarrow$0.0\%) & 300.00 ($\downarrow$0.0\%) & 37.75 ($\downarrow$0.0\%)           & 5.36 ($\uparrow$1.5\%)           & 300.00 ($\downarrow$0.0\%)          & 55.41 ($\downarrow$49.5\%)           & 22.74 ($\downarrow$59.6\%)         & 230.99 ($\downarrow$23.0\%) \\
\vanillafour                    & 67.10 ($\downarrow$36.0\%)          & 10.60 ($\downarrow$0.0\%) & 300.00 ($\downarrow$0.0\%) & 37.77 ($\uparrow$0.0\%)           & 4.92 ($\downarrow$6.8\%)           & 300.00 ($\downarrow$0.0\%)          & 60.86 ($\downarrow$44.5\%)           & 20.06 ($\downarrow$64.4\%)         & 300.00 ($\downarrow$0.0\%)           \\
\hline
\rbotthree                     & \textbf{55.71 ($\downarrow$46.9\%)}          & 10.41 ($\downarrow$1.8\%) & 300.00 ($\downarrow$0.0\%) &  26.19 ($\downarrow$30.6\%) & 4.61 ($\downarrow$12.7\%) & 35.25 ($\downarrow$88.2\%) & 37.71 ($\downarrow$65.6\%)           & 8.37 ($\downarrow$85.1\%)         & 65.67 ($\downarrow$78.1\%) \\
\rbotfour                       & 57.60 ($\downarrow$45.1\%) & \textbf{10.37 ($\downarrow$2.2\%)} & 300.00 ($\downarrow$0.0\%) & \textbf{25.35 ($\downarrow$32.9\%)} & \textbf{4.58 ($\downarrow$13.2\%)} & \textbf{17.17 ($\downarrow$94.3\%)} & \textbf{12.45 ($\downarrow$88.6\%)} & \textbf{5.04 ($\downarrow$91.0\%)} & \textbf{48.30 ($\downarrow$83.9\%)} \\ \hline
\end{tabular}
}
\vspace{-1.5em}
\end{table*}

\hi{Step 2: Recipe-based Rule Selection and Ordering.} 
Derived recipes offer more detailed guidance for rewriting the input query compared to the basic rule specifications and \QAs. Therefore, we refine the initially arranged rule sequence to align it more closely with these recipes. This process is divided into three sub-steps.

$(i)$ We first refine rule selection with recipes. A straightforward method is to assess the relevance between each rule and each recipe, excluding those rules deemed irrelevant. However, this method faces two significant drawbacks. Firstly, it overlooks rule dependencies, consequently excluding indirectly utilized rules that become relevant only after another rule is applied. Secondly, this method incurs high costs, especially since it lies on the critical path for online query rewrite. To address these limitations, we propose a batch filtering method. Specifically, we first feed the rewrite recipe and a batch of rules to \llm, and require \llm to select the most appropriate rules aligned with the rewrite recipe, using the prompt \textit{$p_{select\_rule}=$``Given the input query and rewrite rules, you should evaluate whether the rules can be applied to rewrite the query in the context of the recipe.''}  Next, we feed the previously selected rules along with the next batch of rules into \llm for further selection. We continue this process iteratively until all rules have been evaluated. In this way, we not only accelerate the \llm selection process, but also consider the rule dependency within the selected rules.

$(ii)$ Next, we methodically refine the rule order according to rewrite recipe. First, taking into account the inter-dependence of the selected rules, we categorize them into groups wherein each group pertains to the same SQL operator (e.g., join). We then instruct \llm to arrange them following the rewrite recipe, emphasizing the alignment of closely related rules. Second, leveraging the grouped rule ordering as a basis, we instruct \llm to refine the overall rule ordering to optimally align with the rewrite recipe. 

$(iii)$ We input the refined rule sequence to the query rewrite engine, and rewrite the input query with the rules.

\hi{Step 3: Rewrite Reflection.}
With regards to the unstable performance of \llm in complex tasks like rule arrangement~\cite{wei2022chain, yao2024tree}, we reflect the rewrite process to determine whether to finalize or refine the rewrite as discussed in Section~\ref{sec:overview}.

\vspace{-.5em}

\section{Experiments}
\label{sec:experiments}

\vspace{-.25em}
\subsection{Experiment Setting}
\label{sec:exp-setting}
\vspace{-.25em}

We implement our system \oursys using the rules in an open-sourced query engine Apache Calcite~\cite{begoli2018apache}. We execute SQL queries in PostgreSQL v14 on a machine with 128 GB RAM and 3.1GHz CPU.

\vspace{-.15em}
\hi{Datasets.}
To verify the effectiveness of \oursys on different scenarios, we conduct experiments on three types of datasets. 
$(i)$ TPC-H is a standard OLAP benchmark, which contains 62 columns and 44 queries. 
We separately test \oursys on different data sizes, i.e., TPC-H 10x ($\sim$10G) and TPC-H 50x ($\sim$50G). 
$(ii)$ DSB is a more complex OLAP benchmark {adapted from TPC-DS}~\cite{ding2021dsb}, which contains 429 columns and 76 queries. 
We also test \oursys on datasets DSB 10x ($\sim$10G) and DSB 50x ($\sim$50G).
$(iii)$ Calcite is constructed from Apache Calcite's rewrite rule test suites~\cite{calcite}, which is widely used in evaluating query rewrite capabilities~\cite{wang2022wetune,dong2023slabcity,bai2023querybooster}. 
In our experiment, we select 44 representative queries with great rewrite potentials on a schema of 43 columns.~\footnote{For efficiency, we chose Calcite test queries with top 11\% performance gains under Calcite rewrite. Additional experiments on full queries confirmed that \oursys consistently reduces query latency and outperforms baseline methods.} Additionally, we evaluate \oursys with a 10G data size across distinct data distributions, namely Calcite (uni) for uniform distribution and Calcite (zipf) for Zipfian distribution.

\begin{figure}[!t]
	\centering
\includegraphics[width=.8\linewidth]{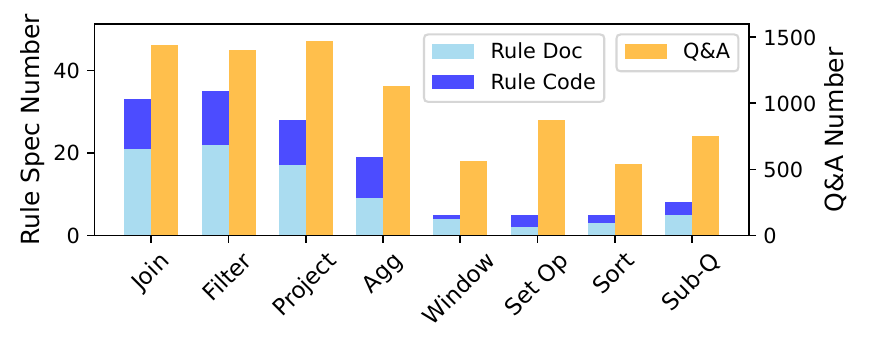}
\vspace{-2em}
	\caption{{Distribution of Rewrite Evidences.}}
	\label{fig:rewrite-dataset-dist}
\vspace{-1.5em}
\end{figure}

\begin{figure}[!t]
	\centering
\includegraphics[width=\linewidth]{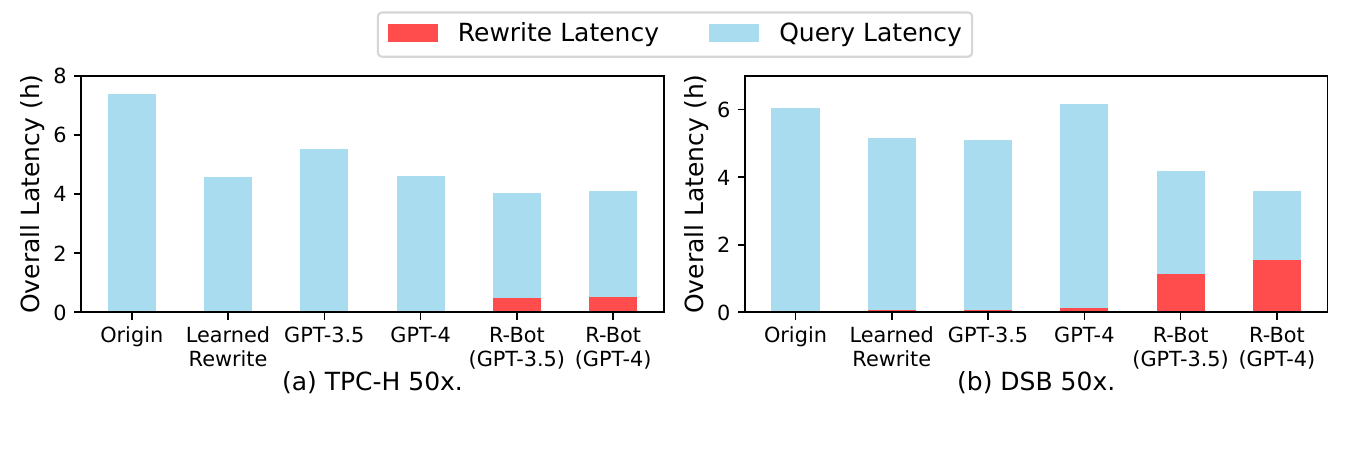}
\vspace{-3.5em}
	\caption{{Comparison of Overall Latency.}}
	\label{fig:overall-latency}
\vspace{-2em} 
\end{figure}

\vspace{-.15em}
\hi{Rewrite Evidences.}
Our prepared rewrite evidences are shown in \autoref{fig:rewrite-dataset-dist}, including 67 expert-verified rewrite rule specifications (30 from 80k+ tokens of documents, 37 from 30k+ lines of code), and 2091 filtered rewrite \QAs from millions of website Q\&As. 
Together, they cover diverse SQL operations (e.g., joins, sub-queries) and support a wide range of rewrite scenarios.

\vspace{-.15em}
\hi{LLMs.}
We use the state-of-the-art \llm gpt-4o, and alternatively cheaper gpt-3.5-turbo-0125~\cite{openai}. We use the default temperature of 0.1, in balance of reproduction and \llm performance~\cite{llamaindex}.

\vspace{-.15em}
\hi{Evaluation Metrics.}
We evaluate \oursys using two metrics.
$(i)$ Query Latency,  which measures the execution time of the rewritten query. 
$(ii)$ Overall Latency, which captures the total time spent on both rewriting and executing the query. 
For each query, we conduct five executions and calculate the average, excluding the highest and lowest values. 
For each metric, we evaluate performance based on the average, median, and 90th percentile (p90).

\vspace{-.15em}
\hi{Evaluated Methods.} 
The evaluated methods include:
$(i)$ \rbotfour is \oursys using gpt-4o.
$(ii)$ \rbotthree is \oursys using gpt-3.5-turbo-0125.
$(iii)$ \lr employs Monte Carlo Tree Search (MCTS) to explore the optimal sequence of rewrite rules~\cite{zhou2021learned}. It estimates the performance improvements of the rewritten queries with {query cost models}, which guides the search process. ~\footnote{{Learning-based query rewrite methods, including LLM-based approaches that require model training~\cite{lillm}, are generally unsuitable for our tests due to the absence of a train-test dataset split. As for \lr, we use the authoritative version~\cite{lrrepo}, integrating query cost models approximated using query plan statistics.}}
$(iv)$ \vanillafour without using techniques in \oursys, which inputs the SQL query and the rewrite rules, and outputs the arranged rule sequence.
$(v)$ \vanillathree similarly without using techniques in \oursys.

\vspace{-.75em}
\subsection{Performance Comparison}
\label{sec:perf-comp}
\vspace{-.25em}

We compare \oursys with two types of baselines, including learning-based methods (\lr), and origin \llms (\vanillafour, \vanillathree).

\hi{Query Latency Reduction.}
\autoref{tab:methods-execution-time} shows the query latency of rewritten queries. \oursys outperforms the other query rewrite methods across all the datasets and metrics. The reasons are two-fold.

First, \oursys can judiciously retrieve relevant rule specifications and \QAs to figure out the rewrite rules pertinent to the input query.
For instance, \oursys can apply the exploration rule ``AGGREGATE\_EXPAND\_DISTINCT\_AGGREGATES\_TO\_JOIN'' weakly related to the input query, using retrieved evidences to achieve a 5.6x optimization.
However, the other methods fail to identify this critical rule without guidance from rewrite evidences. 
Besides, we find that \vanillafour and \vanillathree tend to select only a small number of rules (e.g., 1), which is often sub-optimal.
That is because, they {exhibit hallucination without rewrite evidences}, and thus miss the intricate correlations between rewrite rules and the input query.

Second, \oursys adopts a step-by-step \llm rewrite method, which can leverage rewrite evidences to understand the interrelations among the rewrite rules during rule arrangement.
For instance, \oursys can accurately select and arrange multiple rewrite rules (e.g., 6) from analysis of retrieved evidences, so that they can co-operate together to rewrite the input query to an execution-efficient form (e.g., 4.6x accelerated). On the contrary, since \lr relies on blind exhaustive search, it can hardly find the optimal rule sequence among tremendous possibilities (e.g., only 4.3x accelerated).

\hi{Overall Latency Reduction.}
We also evaluate overall latency on larger TPC-H 50x and DSB 50x datasets.  As shown in \autoref{fig:overall-latency}, \oursys achieves the best latency, demonstrating improvements of 1.82x and 1.68x respectively. The reasons are three-fold. First, \oursys can still optimize the query latency by finding better plans, thus outperforming other methods. Second, \oursys takes some time (e.g., average around 1 min) to retrieve beneficial evidences and iteratively instruct \llm to select and arrange rewrite rules. Thus it can steadily identify the critical rewrite rules (e.g., ``FILTER\_SUB\_QUERY\_TO\_JOIN'') with much higher query latency reduction.
Instead, other methods do not spend efforts understanding rewrite evidences, which often causes sub-optimal rewritten queries.
Third, compared with results of TPC-H 10x and DSB 10x (see \autoref{tab:methods-execution-time}), we find that \oursys remains similar rewrite latency but demonstrates higher query latency reduction proportional to data scale, which both results in superior overall latency.

\begin{table}[!t]
\centering
\vspace{-2em}
\caption{Comparison of Query Improvement Ratio.}
\label{tab:methods-improvement}
\vspace{-1em}
\hspace*{-1em}\begin{tabular}{cccc}
\hline
Improve (\#Queries) & TPC-H 10x                                                             & DSB 10x                                                               & Calcite (uni)                                                           \\ \hline
\lr & 7/44 (15.9\%) & 4/76 (5.3\%) & 29/44 (63.6\%) \\
\vanillathree                                               & 3/44 (6.8\%)            & 4/76 (5.3\%)            &  16/44 (36.4\%) \\
\vanillafour                                                  & 6/44 (13.6\%)            & 4/76 (5.3\%)            & 21/44 (47.7\%)           \\
\hline
\rbotthree                                                   & \textbf{21/44 (47.7\%)} & 16/76 (21.0\%) & {31/44 (70.4\%)} \\
\rbotfour                                                    & 17/44 (38.6\%)          & \textbf{18/76 (23.7\%)} & \textbf{39/44 (88.6\%)} \\ \hline
\end{tabular}
\end{table}

\begin{table}[!t]
\centering
\vspace{-1em}
\caption{Comparison of Query Latency on Calcite (zipf).}
\label{tab:latency-zipf}
\vspace{-1em}
\hspace*{-1em}
\begin{tabular}{cccc}
\hline

Query Latency (s) & Average        & Median                                                     & p90 \\ \hline
Origin                              & 106.31                                                            & 37.91                                                  & 300.00 \\
\hline
\lr & 71.24 ($\downarrow$33.0\%) &           {5.04 ($\downarrow$86.7\%)} & 300.00 ($\downarrow$0.0\%)                                 \\
\vanillathree                  & 58.33 ($\downarrow$45.1\%)          & 20.04 ($\downarrow$47.1\%) & 300.00 ($\downarrow$0.0\%) \\
\vanillafour                    & 61.80 ($\uparrow$41.9\%)          & 14.15 ($\downarrow$62.7\%) & 300.00 ($\downarrow$0.0\%)           \\
\hline
\rbotthree                       & 32.44 ($\downarrow$69.5\%)          & 6.58 ($\downarrow$82.6\%) & 57.40 ($\downarrow$80.9\%)   \\ 
\rbotfour                     & \textbf{7.56 ($\downarrow$92.9\%)}          & \textbf{4.96 ($\downarrow$86.9\%)} & \textbf{18.08 ($\downarrow$94.0\%)}  \\
\hline
\end{tabular}
\vspace{-1.75em}
\end{table}

\hi{Query Improvement Ratio.} \autoref{tab:methods-improvement} shows the query-level latency reduction ratio by different query rewrite methods, with \oursys ranging from about 21.0\% to 88.6\%. We find that \oursys still outperforms the other methods on the three datasets. That is because \oursys can discover targeted beneficial rule arrangements using relevant evidences, while the other methods have difficulty specifying the optimal arrangement among tremendous search space.

\hi{{Zero-Shot} Robustness.} First, as shown in \autoref{tab:methods-execution-time}, \oursys outperforms the other query write methods on different datasets without any re-training. This is because the generalizability of pre-trained \llm enables \oursys to automatically adapt to unseen database schema and workload. Besides, \autoref{tab:latency-zipf} further demonstrates the robustness of query rewrite methods on data distribution, which transfers the Calcite dataset from uniform distribution to zipf distribution. Compared with Calcite (uni) (see \autoref{tab:methods-execution-time}), we find that the performance of \oursys remains consistent.  \oursys still achieves the lowest average, median, and p90 of query latency among the methods. 
That is because the rewrite evidences can cover different query rewrite scenarios, and \oursys can adaptively decide among potential rule arrangements with feedbacks of database query cost. 

\hi{Evaluation Across Various \llms.}
As shown in \autoref{tab:methods-execution-time}, on most metrics, \rbotthree performs worse than \rbotfour, but still outperforms the other query rewrite methods.
That is because \oursys decomposes complex tasks into simpler stages, both in evidence retrieval and step-by-step \llm rewrite, where each stage can be still manageable even with less advanced \llms like GPT-3.5.

\vspace{-.5em}
\subsection{Ablation Study}
\label{sec:ablation}
\vspace{-.25em}

\subsubsection{Rewrite Evidence Preparation.} 
We evaluate the essence of rewrite evidence by comparing \vanillafour with naive RAG.
As shown in \autoref{tab:ablation-study-evidence}, naive RAG equipped with evidences can outperform \vanillafour across all the metrics, because naive RAG can retrieve rewrite evidences that are somewhat relevant to the input query, which provides beneficial guidance in rewrite rule selection and ordering.

\vspace{-.25em}
\subsubsection{Structure-Semantics Evidence Retrieval.}
To evaluate the effectiveness of hybrid structure-semantics \QA retrieval, we evaluate \oursys separately with semantics-only and structure-only retrieval. As indicated in \autoref{tab:ablation-study}
, both methods exhibit a decline in performance compared to \oursys. 
This reduction is due to their inability to identify \QAs beneficial for query rewrite from a holistic perspective, as they either overlook query semantics or structure information. 
For example, compared with \oursys, they may miss key evidence, leading to the omission of critical rules like ``AGGREGATE\_EXPAND\_DISTINCT\_AGGREGATES'' in arrangement.

\begin{table}[!t]
\centering
\vspace{-2em}
\caption{Ablation Study of Rewrite Evidence on Calcite (uni).}
\vspace{-1em}
\label{tab:ablation-study-evidence}
\begin{tabular}{cccc}
\hline
Query Latency (s) & Average                                                              & Median                                                     & p90 \\ \hline
Naive RAG & \textbf{39.13} & \textbf{14.88} & \textbf{98.15} \\
\vanillafour & 60.86 & 20.06 & 300.00 \\ \hline
\end{tabular}
\end{table}

\begin{table}[!t]
\centering
\vspace{-1em}
\caption{Ablation Study of Evidence Retrieval on Calcite (uni).}
\vspace{-1em}
\label{tab:ablation-study}
\begin{tabular}{cccc}
\hline
Query Latency (s) & Average                                                              & Median                                                     & p90 \\ \hline
Structure-Semantics Retrieval                              & \textbf{12.45}                                                            & \textbf{5.04}                                                  & \textbf{48.30} \\
Structure-Only Retrieval & 31.96 & 5.30 & 56.40 \\
Semantics-Only Retrieval & 39.45 & 8.37 & 65.67 \\
Naive RAG & 39.13 & 14.88 & 98.15 \\ \hline
\end{tabular}
\vspace{-1.25em}
\end{table}

\begin{figure}[!t]
	\begin{minipage}{0.48\linewidth}
	\centering
\includegraphics[width=\linewidth]{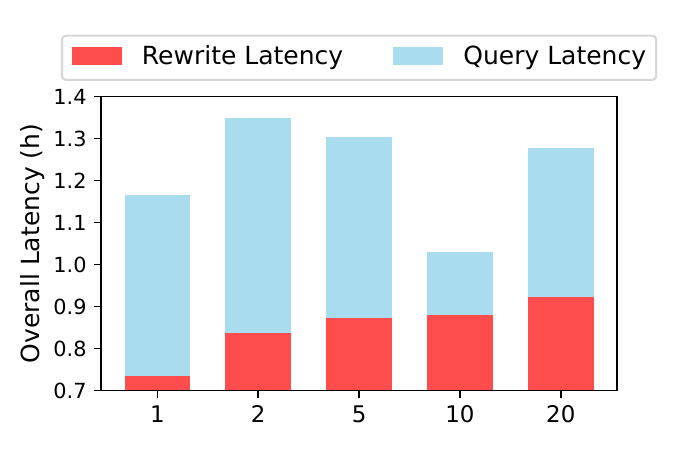}
\vspace{-3em}
\caption{Impact of Retrieval Top-k on Calcite (uni).}
	\label{fig:topk}
	\end{minipage}
	\hfill
	\begin{minipage}{0.48\linewidth}
	\includegraphics[width=\linewidth]{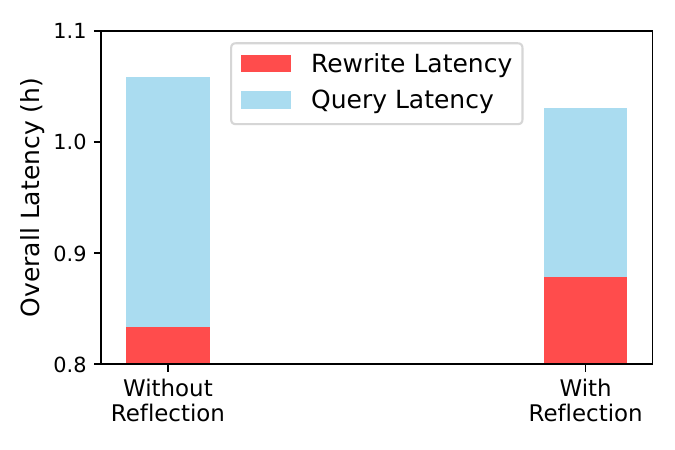}
	\vspace{-3em}
 \caption{Impact of Rewrite Reflection on Calcite (uni).}
	\label{fig:reflection}
	\end{minipage}
  \vspace{-1.25em}
\end{figure}

\begin{table}[!t]
\centering
\caption{Ablation Study of \llm Rewrite on Calcite (uni).}
\vspace{-1em}
\label{tab:ablation-study-steps}
\begin{tabular}{cccc}
\hline
Query Latency (s) & Average                                                              & Median                                                     & p90 \\ \hline
Step-by-Step \llm Rewrite                              & \textbf{12.45}                                                            & \textbf{5.04}                                                  & \textbf{48.30} \\
One-Step \llm Rewrite & 36.47 & 13.83 & 69.43 \\ \hline
\end{tabular}
\vspace{-1em}
\end{table}

\hi{Retrieval Top-k.}
We also evaluate the impact of retrieval top-$k$, which decides the number of retrieved \QAs.
As shown in \autoref{fig:topk}, we have two observations.
First, rewrite latency grows with $k$, because \oursys tends to generate more recipes with a greater number of retrieved \QAs. Consequently, as the recipes compose longer \llm contexts, each invocation of \llm requires additional time to process the extended context.
Second, query latency decreases when $k$ increases from 1 to 10, but rises again beyond 10.
The reasons are two-fold. $(i)$ When $k$ is small, the additional retrieved \QAs are more likely to cover essential evidence, allowing \oursys to select critical rules (e.g., ``AGGREGATE\_JOIN\_TRANSPOSE''). 
$(ii)$ For large $k$, it is likely to already cover the essential evidence in the retrieved \QAs, making further increase unnecessary as it may not provide additional information.
Besides, increasing $k$ too much may introduce noisy contexts, which may impair performance of \llm in query rewrite.
Therefore, we set $k=10$ for the other experiments.
Meanwhile, we achieve a Precision@10 of 19.8\%, indicating that on average at least one relevant \QA is successfully retrieved.

\vspace{-.25em}
\subsubsection{Step-by-Step \llm Rewrite.}
We compare our step-by-step \llm rewrite with one-step \llm rewrite that instructs \llm to arrange the rules in one step using retrieved evidences, and evaluate the performance. As shown in \autoref{tab:ablation-study-steps}, we find that one-step \llm rewrite often selects a smaller number of rules, which results in sub-optimal rule arrangements. That is because, given the extremely long context of retrieved evidences and available rewrite rules, \llm tends to produce short-cut solutions overlooking many critical details, and thus only selects the rules explicitly related to the input query. Instead, \oursys employs a step-by-step \llm rewrite for reasoning, enabling the selection of higher-quality rules.

\vspace{-.25em}
\subsubsection{Rewrite Reflection.}
We also evaluate the impact of rewrite reflection in \oursys.
As shown in \autoref{fig:reflection}, we have two observations.
First, the rewrite latency increases due to the additional overhead introduced by rewrite reflection and \llm self-correction. Second, the query latency decreases with further rewrite reflection, leading to an overall latency reduction. That is because rewrite reflection helps mitigate the hallucination issue in \llms by exploring multiple rewrite rule arrangements to find the most beneficial query rewrite.

\begin{figure}[!t]
	\centering\vspace{-2em}
\includegraphics[width=1.05\linewidth]{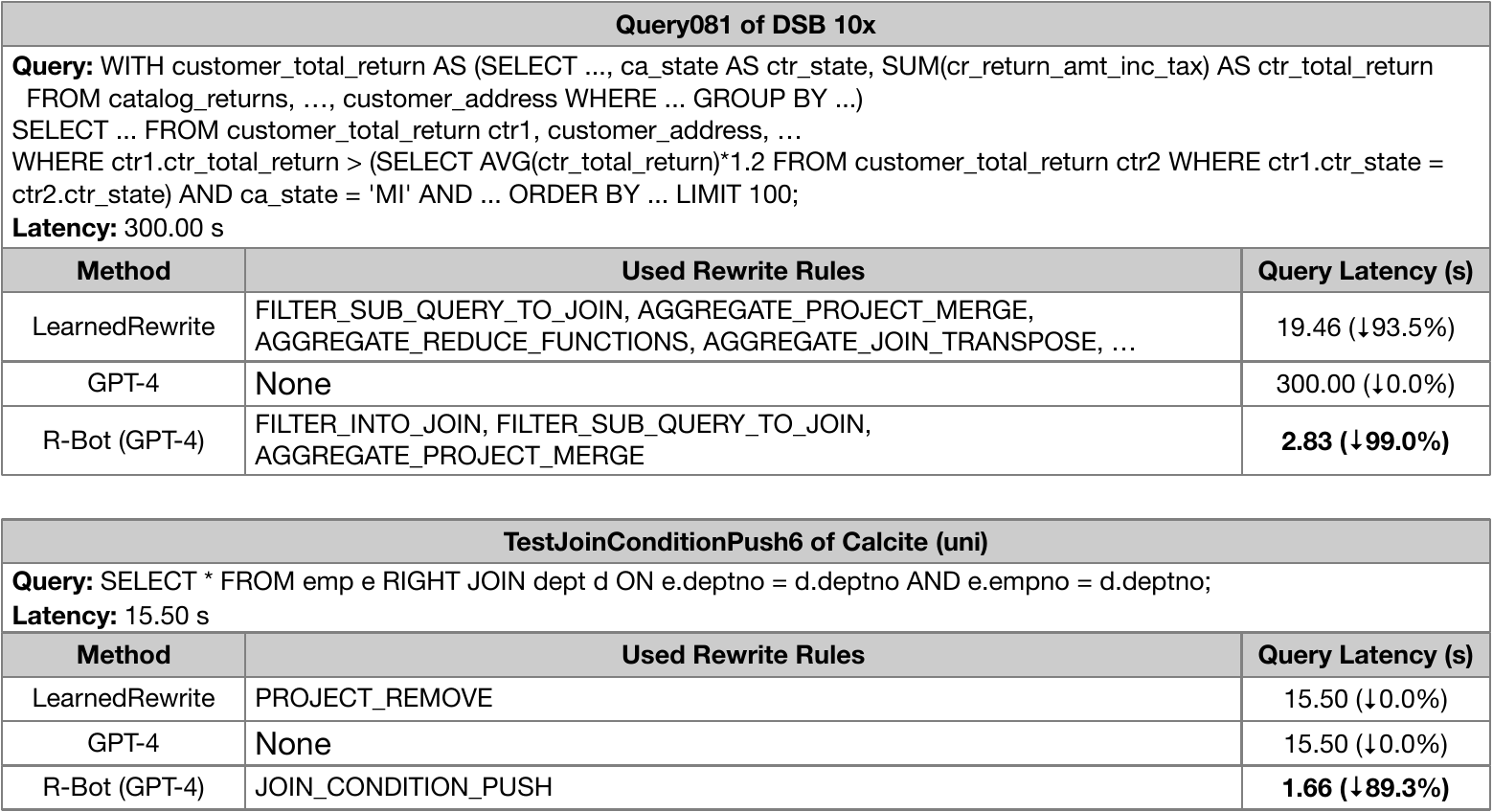}
\vspace{-2em}
	\caption{{Example Query Rewrite Results.}}
	\label{fig:case-study}
\vspace{-1em}
\end{figure}

\vspace{-.75em}
\subsection{Case Study}
\vspace{-.25em}
We provide some representative examples to illustrate the query rewrite results. As shown in \autoref{fig:case-study}, \oursys outperforms other methods by accurately identifying critical rules ``FILTER\_INTO\_JOIN'' and  ``FILTER\_SUB\_QUERY\_TO\_JOIN''  in the first case study, which pushes down conditions involving only one join table and removes the time-consuming sub-queries in the query. On the contrary, \vanillafour selects no useful rewrite rules due to absence of rewrite evidences and dedicated \llm algorithm design. Besides, \lr tends to select an excessive number of rewrite rules under erroneous guidance, leading to a less efficient rewritten query. In the second case study, \oursys also demonstrates superior performance by selecting the rewrite rule ``JOIN\_CONDITION\_PUSH'', which derives an additional condition ``$e.deptno=e.empno$''. This condition can be pushed down into the sub-query involving the table ``$emp$'', a refinement overlooked by other query rewrite methods. However other methods cannot detect this rule.

\vspace{-.75em}
\subsection{Deployment at Huawei with Real Customers}
\label{sec:deploy}
\vspace{-.25em}

\begin{figure}[!t]
	\centering\vspace{-2em}
\includegraphics[width=.95\linewidth]{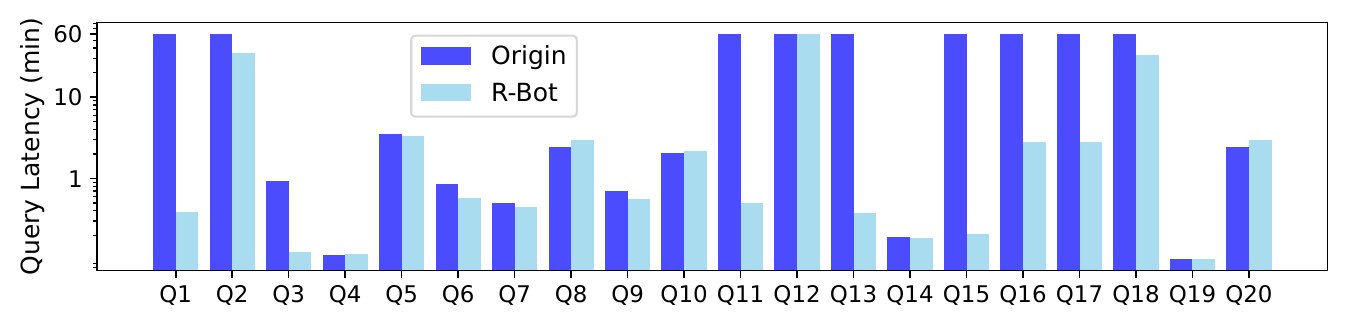}
\vspace{-1em}
	\caption{{Performance of \oursys on Real Dataset.}}
	\label{fig:real-workload}
    \vspace{-1.5em}
\end{figure}

We have also deployed \oursys at Huawei, which is empowered by open-sourced \llm DeepSeek-R1-Distill-Qwen-32B~\cite{deepseekr1} and embedding model gte-Qwen2-1.5B-instruct~\cite{gteqwen2}.
We verify its effectiveness using a real-world dataset on the largest bank in China, containing 190 columns and 74GB data, as well as 20 real-world slow queries with 7 joined tables and 3 sub-queries on average.
As shown in \autoref{fig:real-workload}, we find that \oursys can effectively optimize these queries, i.e., improving the latency of 70\% queries and reducing the overall latency from 9.23 hours to 4.37 hours. Notably, \oursys remains effective even when $(i)$ the tested queries are absent from the \llm pre-training corpus (not publicly available) and $(ii)$ open-sourced 32B \llm and embedding model are used as the backend.

\vspace{-.5em}
\section{Related Work}
\vspace{-.25em}

\hi{Query Rewrite.}
Existing query rewrite methods mainly adopt two paradigms.
(1) Heuristic-based methods~\cite{mysql,postgresql,graefe1993volcano,graefe1995cascades,begoli2018apache}.
Some methods (e.g., PostgreSQL~\cite{postgresql}) apply rewrite rules in a fixed order, which often overlook better orders in different scenarios.
Besides, other heuristic-based methods (e.g., Volcano~\cite{graefe1993volcano}) attempt to explore different rule orders with the aid of heuristic acceleration. However, by neglecting inter-dependencies among the rules, these methods perform a blind search, often failing to identify the optimal orders within reasonable time.
(2) Learning-based methods~\cite{zhou2021learned,lillm}.
\lr decides the rule order using Monte Carlo Tree Search guided by learned cost models.
However, the learned model cannot be transferred to unseen database schema without additional retraining, which is often impractical.
Hence, it calls for a query rewrite system capable of reliably identifying an appropriate sequence of rewrite rules to optimize query rewrite. 
A recent \llm-based query rewrite method addresses hallucination by incorporating query rewrite examples into the context for \llm to emulate~\cite{lillm}. However, it still lacks robustness, as it depends on an example pool generated from the training dataset and utilizes a trained model to select examples based on SQL query similarity.

\hi{Automatic Rewrite Rule Discovery.}
Given the tedious effort required to curate a large number of high-quality rewrite rules, some methods have been recently proposed to automate the process of rule discovery~\cite{wang2022wetune,dong2023slabcity,ding2023proving}. \textit{WeTune}~\cite{wang2022wetune} first represents the condition and transformation of rewrite rule with operators and symbols. Then it enumerates potential rewrite rules and verifies rule equivalence using SMT-solvers~\cite{zhou2022spes,chu2017cosette}. \textit{SlabCity}~\cite{dong2023slabcity} discovers possible query rewrites for a specific query by recursively applying various transformations to the original query.  
Complementary to the above works, \oursys retrieves highly relevant evidences to instruct \llm, effectively arranges newly discovered rules, and achieves superior adaptability and robustness in rule selection and ordering.

\hi{\llm for Database.}
There are also many studies to use LLMs to optimize databases~\cite{achiam2023gpt,liu2024survey,fan2024combining,lao2024gptuner,zhou2024d,sun2025dbot,zhou2024db,liu2024query,zhou2025survey}. For instance, \textit{GPTuner}~\cite{lao2024gptuner} enhances database knob tuning using \llm by leveraging domain knowledge to identify and tune knobs. \textit{D-Bot}~\cite{zhou2024d,sun2025dbot} is an \llm-based database diagnosis system. 
Unlike these approaches, \oursys is a novel query rewrite system powered by \llm.  

\vspace{-.25em}
\section{Conclusion}
\vspace{-.25em}

We proposed an \llm-based query rewrite system.
First, we prepared rewrite evidences from diverse sources, including rewrite rule specifications and  rewrite \QAs.
Next, we proposed a hybrid structure-semantics retrieval method to retrieve relevant rewrite evidences, based on which we generated rewrite recipes to instruct \llm for query rewrite.
Then, we proposed a step-by-step \llm method, which iteratively utilized the retrieved \QAs and rewrite recipes to select and arrange rewrite rules with self-reflection.
Experimental results demonstrated that \oursys achieved remarkable improvements over existing  methods. Moreover, \oursys deployed at Huawei and with real customers showed the effectiveness of \oursys.

\begin{acks}
This paper was supported by National Key R\&D Program of China (2023YFB4503600),
NSF of China (62525202, 62232009), Shenzhen Project (CJGJZD20230724093403007), Zhongguancun
Lab, Huawei, and Beijing National Research Center for Information Science and Technology
(BNRist). Guoliang Li is the corresponding author.
\end{acks}

\clearpage

\balance
\bibliographystyle{ACM-Reference-Format}
\bibliography{refs}
\balance

\end{document}